\documentclass[%
 reprint,
 amsmath,amssymb,
 aps,
]{revtex4-2}

\usepackage{graphicx}
\usepackage{dcolumn}
\usepackage{bm}
\usepackage{bm}
\usepackage{xcolor}
\newcommand{\norm}[1]{\left\lVert#1\right\rVert}
\newcommand{\comment}[1]{}
\newcommand{\nb}{n_{\text{b}}}
\newcommand{\ee}[1]{\text{e}^{#1}}
\newcommand{\GV}{\widehat{\text{GV}}}
\newcommand{\hG}{\hat{\text{G}}}
\allowdisplaybreaks

\begin{document}

\preprint{APS/123-QED}

\title{Unified treatment of exact and approximate scalar electromagnetic wave scattering}

\author{Subeen Pang}
\email{sbpang@mit.edu}
\author{George Barbastathis}
\altaffiliation[Also at: ]{Singapore-MIT Alliance for Research and Technology (SMART) Centre, 1 CREATE Way, Singapore 138602, Singapore.}
\affiliation{Department of Mechanical Engineering, Massachusetts Institute of Technology, Cambridge, Massachusetts 02139, USA}

\date{\today}

\begin{abstract}
Under conditions of strong scattering, a dilemma often arises regarding the best numerical method to use. Main competitors are the Born series, the Beam Propagation Method, and direct solution of the Lippmann-Schwinger equation. However, analytical relationships between the three methods have not yet, to our knowledge, been explicitly stated. Here, we bridge this gap in the literature. In addition to overall insight about aspects of optical scattering that are best numerically captured by each method, our approach allows us to derive approximate error bounds to be expected under various scattering conditions.
\end{abstract}

\maketitle

\section{Introduction}
In computational imaging, quantitative physical properties of objects are estimated from optical measurements of scattered fields. The complex light-matter interactions leading to scattering are governed by Maxwell's equations or, under some assumptions, by the scalar Helmholtz equation that describes optical elastic scattering from objects that are large compared to the wavelength \cite{paganin2006coherent}. 

To simplify the process of modeling optical scattering and estimating object properties, there have been many studies on approximating solutions to the scalar Helmholtz equation. One of the most primitive is the projection approximation, where the scattered field is assumed to maintain the incidenct wavefront, e.g. a plane or spherical wave, while attenuation and phase delay accumulate proportional to the optical path length of rays through the object. This assumption leads to the Radon transform formulation, and is the basis of computed tomography. A more elaborate description is provided by the so-called single scattering approximations, including the first Born and Rytov methods \cite{marks2006family}. As objects become dense and highly scattering, as expected, even single scattering methods start to fail, and models accounting for multiple scattering are required. Representative approaches are the Lippmann-Schwinger equation (LSE) \cite{pham2020three,burgel2017sparsity,liu2017seagle}, the beam propagation method (BPM) \cite{kamilov2015learning,kamilov2016optical,goy2019high} and the Born series \cite{osnabrugge2016convergent,tahir2019holographic}.

Multiply scattering models can all be formulated starting from the scalar Helmholtz equation, but they rely on different approximations on the scattering process \cite{kruger2017solution, hohage2006fast, feit1988beam, paganin2006coherent, colton1998inverse}. Subsequently, all three aforementioned methods may exhibit certain drawbacks compared to exact solutions of the scalar Helmholtz equation, and the discrepancies evidence themselves differently for each method. For example, it has been reported that BPM cannot account for backscattering or reflection of fields and it would not be suitable for experimental conditions that significantly deviate from the paraxial approximation \cite{lim2019high,chen2020multi}. Born series is numerically unstable, unless the optical potential is sufficiently weak. On the contrary, the LSE, by virtue of originating simply as an integral formulation of the scalar Helmholtz equation under the standard Rayleigh-Sommerfeld radiation condition, requires no further assumptions. In principle, this can lead to high-precision solutions in numerically ideal cases \cite{colton1998inverse,hohage2006fast,ying2015sparsifying}. However, solving the LSE may still be subject to numerical artifacts resulting from the inversion of the integral equation, and requires relatively intensive computational resources.

Hence, while the LSE promises the most reliable approximations of scattered fields and optical objects \cite{pham2020three}, we can consider using BPM or Born series if an error compared to LSE is bounded below a given acceptable threshold. In previous studies, conditions that can make such small error achievable are usually summarized qualitatively, e.g. laterally large objects, small illumination angles, and weak potential. This is because LSE, BPM, and Born series originate from different approximations and derivations. Subsequently, explicit and quantitative relationships between the different methods, especially between LSE and BPM, have not been addressed very clearly. 

In fact, the precision of a scattering model may not be the sole parameter to determine the quality of field/object estimations. This is because such estimations consist of complex optimization procedures, which would also depend on various mathematical conditions e.g. preconditioning and regularization. Nevertheless, a more concrete understanding of the relationships and relative strengths and weaknesses of each method would be beneficial for us to analyze estimation results, review numerical settings, and track origins of artifacts and errors by evaluating applicability of scattering models.

Therefore, in this paper, we propose a definitive and quantifiable relationship among LSE, Born series, and BPM and introduce concrete conditions where the scattered fields estimated respectively from the three methods exhibit insignificant differences. Specifically, we first suggest a dimensionless parameter that is easy to evaluate and can be used to test the validity of Born series solution. Furthermore, we derive the BPM from the LSE and its corresponding Born series. This leads to another dimensionless parameter based on explicit approximations adopted along the derivation. We expect that our study can help analysis not only of field and object estimations but also of scattering models themselves. We expect that our approach can be extended to other models e.g. \cite{brenner1993light,chen2020multi} that are not discussed in this paper but closely relate to LSE, Born series, and BPM.

\section{Formulation of LSE}
When the wavelength of an incident field is smaller than the length scale of the object, the elastic scattering of fields $\psi(\bm{r})$ is governed by the scalar Helmholtz equation,
\begin{equation} \label{eq:helmholtz}
    \left[\nabla^2 + (\nb k_0)^2 \right]\psi(\bm{r}) 
    = 
    -(\nb k_0)^2 
    \left[ \left(\frac{n(\bm{r})}{\nb}\right)^2 -1 \right]
    \psi(\bm{r}).
\end{equation}
Here, $k_0$ is the wavenumber in vacuum, and $\nb$ and $n(\bm{r})$ are the indices of refraction in the background medium and in the (spatially variant) object, respectively. As a reminder, the phase velocities are obtained by dividing the vacuum light speed by the respective indices. Using the Green's function that satisfies the radiation condition \cite{schmalz2010derivation},
\begin{equation} \label{eq:Greenfunction}
    G(\bm{r}-\bm{r}^\prime) =
    \frac{\exp{( i \nb k_0 \norm{\bm{r}-\bm{r}^\prime})}}{4\pi \norm{\bm{r}-\bm{r}^\prime}},
\end{equation}
we may derive an integral formulation identical to Eq.~\ref{eq:helmholtz}, which is the LSE:
\begin{equation} \label{eq:lse}
    \psi(\bm{r}) = 
    \psi_0(\bm{r})
    +
    \int d\bm{r}^\prime \,\,
    G(\bm{r}-\bm{r}^\prime) 
    V(\bm{r}^\prime)
    \psi(\bm{r}^\prime).
\end{equation}
Here, $V(\bm{r})=(\nb k_0)^2 \left[ \left(\frac{n(\bm{r})}{\nb}\right)^2 -1 \right]$ is the optical scattering potential and $\psi_0$ is the incident field. 

\begin{figure}[t]
    \centering
    \includegraphics[width=0.45\textwidth]{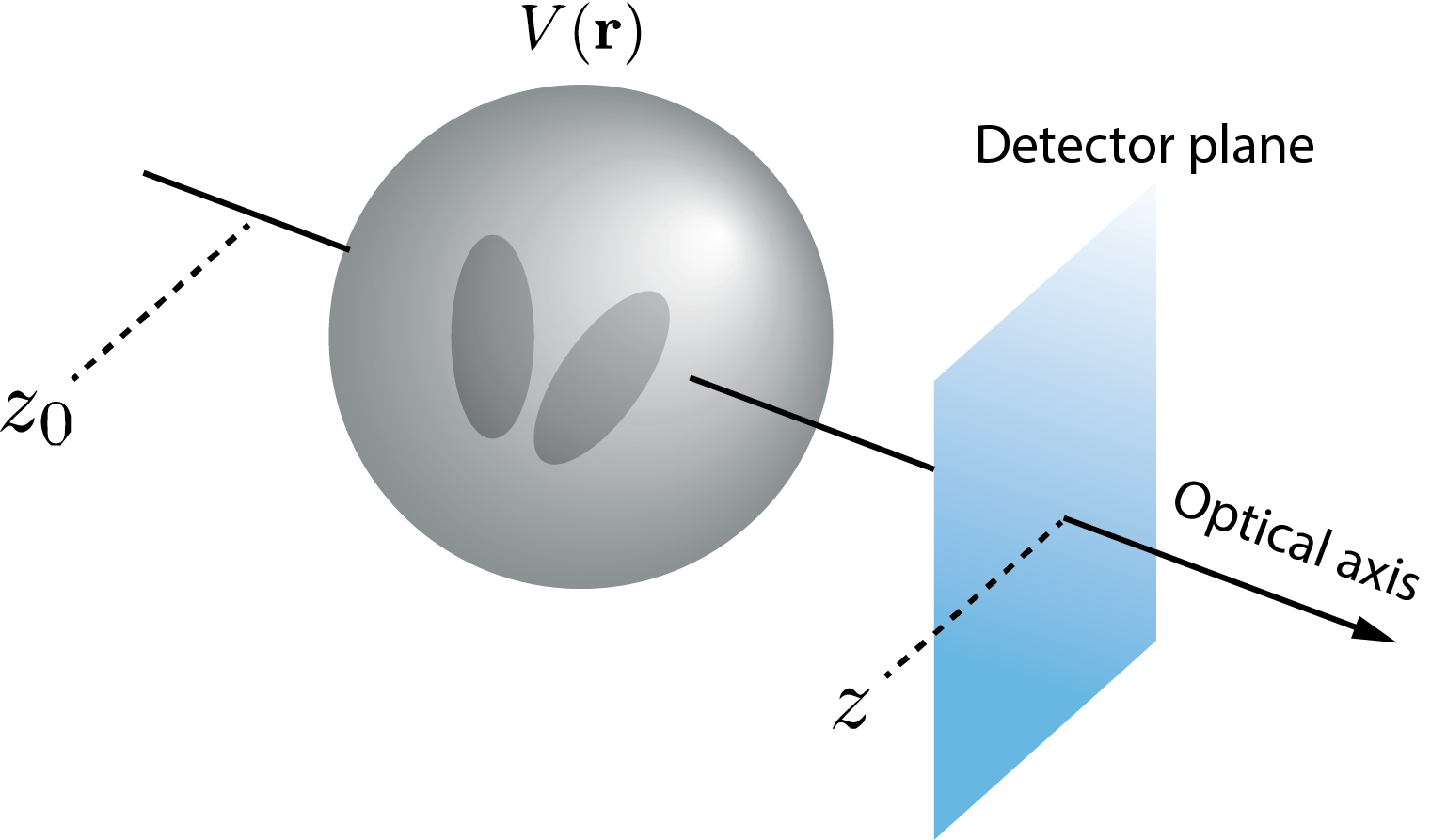}
    \caption{An example geometry for optical scattering from an optical potential $V$.}
    \label{fig:basic_geom}
\end{figure}

The BPM describes the scattering process as a sequential application of 2D scattering layers, so it is not obvious how it can relate to the above LSE development. To develop the relationship later, it will be convenient to re-express the 3D Green's function in terms of its Fourier spectrum. To this end, we use the Weyl expansion \cite{born2013principles} 
\begin{equation} \label{eq:weyl_expansion}
    \frac{\ee{i \nb k_0 r}}{r} = \frac{i}{2\pi} \int dk_x dk_y \,\, \frac{ \ee{i(k_x x + k_y y + k_z |z|)} }{k_z},
\end{equation}
where $r=\norm{\bm{r}}$, $k_z=\sqrt{(\nb k_0)^2 - k_x^2 - k_y^2}$, and $k_x$ and $k_y$ are coordinates in the Fourier space. Setting $z$ to be the optical axis, let us denote $\hat{\mathcal{F}}_{xy}$ as the 2D Fourier transform operator in the lateral dimensions. From the Weyl expansion, the original LSE can be rewritten as a composition of 2D Fourier transforms as 
\begin{align} \label{eq:LSE_2D_FT}
\begin{split}
    &\psi(\bm{r}) - \psi_0(\bm{r}) \\
    &\quad = 
    \frac{i}{2} \int dz^\prime \,\,
    \hat{\mathcal{F}}_{xy}^\dagger
    \left[
    \frac{ \ee{i k_z |z-z^\prime| } }{k_z}\,\,
    \beta (k_x, k_y, z^\prime)
    \right],
\end{split}
\end{align}
where $\dagger$ represents the adjoint operation and 
\begin{equation}
    \beta(k_x, k_y, z) = \hat{\mathcal{F}}_{xy} 
    \left[V(\bm{r})  \psi(\bm{r})\right].
\end{equation}
The full derivation is in Appendix~\ref{app:2DLSE}. Without much loss of generality, we can assume that $\psi_0$ is incident from $z=-\infty$ and the optical detectors are located outside the support of $V$. In addition, let us set $z_0$ as an arbitrary point on the optical axis between the illumination source and the scattering potential $V$. Fig.~\ref{fig:basic_geom} depicts the overall geometry. Consequently, we obtain 
\begin{align} \label{eq:LSE2DOp}
\begin{split}
    &\psi(\bm{r}) - \psi_0(\bm{r}) = \\
    & \qquad =
    \int d\bm{r}^\prime \,
    G(\bm{r}-\bm{r}^\prime) 
    V(\bm{r}^\prime)\psi(\bm{r}^\prime)  \\
    & \qquad =
    \int^{z}_{z_0} dz^\prime \int dx^\prime dy^\prime \,
    G(\bm{r}-\bm{r}^\prime) 
    V(\bm{r}^\prime)\psi(\bm{r}^\prime) \\
    & \qquad =
    \frac{i}{2} \int^z_{z_0} dz^\prime \,\,
    \hat{\mathcal{F}}_{xy}^\dagger
    \left[
    \frac{ \ee{i k_z (z-z^\prime) } }{k_z}
    \beta (k_x, k_y, z^\prime)
    \right],
\end{split}
\end{align}
i.e. the 3D convolution with the Green's function becomes a cascade of 2D convolutions at each $z$-slice.

\section{From LSE to Born series}
To derive a connection between LSE and BPM, we are required to express the original Born series in terms of the cascade of 2D convolutions in Eq.~(\ref{eq:LSE2DOp}). For this, we first slightly modify Eq.~(\ref{eq:LSE2DOp}). Following the small-wavelength approximation underlying the scalar Helmholtz equation or noting that the wavefront envelope of $\psi_0$ would be much larger than objects in many imaging systems, it may be assumed that $\psi_0=\exp(i\nb k_0z)$, {\it i.e.} a pure plane wave. Dividing both sides of Eq.~(\ref{eq:LSE2DOp}) by $\psi_0$, we obtain
\begin{equation} \label{eq:LSEreduced1}
    \varphi(\bm{r}) 
    = 
    1 +
    \frac{i}{2}
    \int^z_{z_0} dz^\prime \,\,
    \hat{\mathcal{F}}_{xy}^\dagger
    \left[
    \frac{ \ee{i \overline{k}_z (z-z^\prime) } }{k_z}
    \gamma (k_x, k_y, z^\prime)
    \right],
\end{equation}
where $\varphi = \psi/\psi_0$, $\overline{k}_z=k_z-\nb k_0$, and
\begin{equation}
    \gamma (k_x, k_y, z) =  \hat{\mathcal{F}}_{xy} 
    \left[V(\bm{r})  \varphi(\bm{r})\right].
\end{equation}
From Eqs.~(\ref{eq:LSE2DOp}) and (\ref{eq:LSEreduced1}), we define an LSE integral operator $\GV_{\alpha}$ as
\begin{align*} \label{eq:LSE2DOp2}
    \GV_{\alpha} :\,\, &\varphi \rightarrow 
    \frac{1}{\psi_0}
    \int^{z}_{\alpha} dz^\prime 
    \int dx^\prime dy^\prime \,\,
    G(\bm{r}-\bm{r}^\prime)  V(\bm{r}^\prime)
    \psi(\bm{r}^\prime) \\
    &=
    \frac{i}{2} \int^z_{\alpha} dz^\prime \,\,
    \hat{\mathcal{F}}_{xy}^\dagger
    \left[
    \frac{ \ee{i \overline{k}_z (z-z^\prime) } }{k_z}
    \gamma (k_x, k_y, z^\prime)
    \right], \stepcounter{equation}\tag{\theequation}
\end{align*}
e.g. $\varphi = 1+\GV_{z_0}\varphi$. In addition, using that $\ee{i \overline{k}_z (z-z^\prime) } = 1$ at the origin of the Fourier space and setting $z_0=-\infty$, we convert Eq.~(\ref{eq:LSEreduced1}) to a more generalized form as
\begin{align*} \label{eq:LSEreduced2}
    &\varphi(\bm{r})
    = 1 + \frac{i}{2} 
    \int^{z_1}_{-\infty} dz^\prime \,\,
    \hat{\mathcal{F}}_{xy}^\dagger
    \left[
    \frac{ \ee{i \overline{k}_z (z-z^\prime) } }{k_z}
    \gamma (k_x, k_y, z^\prime)
    \right]\\
    & \qquad +
    \frac{i}{2} 
    \int^{z}_{z_1} dz^\prime \,\,
    \hat{\mathcal{F}}_{xy}^\dagger
    \left[
    \frac{ \ee{i \overline{k}_z (z-z^\prime) } }{k_z}
    \gamma (k_x, k_y, z^\prime)
    \right] \\
    &= 
    \hat{\mathcal{F}}_{xy}^\dagger
    \ee{i \overline{k}_z (z-z_1) } \hat{\mathcal{F}}_{xy} 
    \Bigg[
    \Bigg. \\
    & \qquad \Bigg. 1+ 
    \frac{i}{2} \int^{z_1}_{-\infty} dz^\prime \,\,
    \hat{\mathcal{F}}_{xy}^\dagger
    \bigg[
    \frac{ \ee{i \overline{k}_z (z_1-z^\prime) } }{k_z}
    \gamma (k_x, k_y, z^\prime)
    \bigg]
    \Bigg] \stepcounter{equation}\tag{\theequation} \\
    & \qquad +
    \frac{i}{2} 
    \int^{z}_{z_1} dz^\prime \,\,
    \hat{\mathcal{F}}_{xy}^\dagger
    \left[
    \frac{ \ee{i \overline{k}_z (z-z^\prime) } }{k_z}
    \gamma (k_x, k_y, z^\prime)
    \right] \\
    &= \hat{\mathcal{F}}_{xy}^\dagger
    \ee{i \overline{k}_z (z-z_1) } \hat{\mathcal{F}}_{xy} 
    \varphi(x,y,z_1) \\
    &\qquad +
    \frac{i}{2} 
    \int^{z}_{z_1} dz^\prime \,\,
    \hat{\mathcal{F}}_{xy}^\dagger
    \left[
    \frac{ \ee{i \overline{k}_z (z-z^\prime) } }{k_z}
    \gamma (k_x, k_y, z^\prime)
    \right],
\end{align*}
where $z_1 \leq z$ is a point on the optical axis. 

Assuming that the operator norm of $\GV_{z_0}$ is less than 1, the solution of the Fredholm integral equation of the second kind, Eq.~(\ref{eq:LSEreduced1}), can be described as a convergent geometric series (Born series or Liouville-Neumann series) \cite{ishizuka1977new}:
\begin{equation} \label{eq:LNseries}
    \varphi(\bm{r}) = \sum_{j=0}^{\infty}
    \left(\frac{i}{2}\right)^j f_j(\bm{r}),
\end{equation}
where
\begin{subequations} \label{eq:fterms}
\begin{equation} \label{eq:f0}
    f_0(\bm{r}) = \hat{\mathcal{F}}_{xy}^\dagger
    \ee{i \overline{k}_z (z-z_0) } 
    \hat{\mathcal{F}}_{xy} 
    \varphi(x,y,z_0)
\end{equation}
\begin{align*} \label{eq:fn}
    f_j(\bm{r}) 
    &= 
    \int^{z}_{z_0} dz^\prime \,\,
    \hat{\mathcal{F}}_{xy}^\dagger
    \frac{ \ee{i \overline{k}_z (z-z^\prime) }  }{k_z}
    \hat{\mathcal{F}}_{xy} 
    \left[V(\bm{r}^\prime) f_{j-1}(\bm{r}^\prime)\right] \\
    &= \frac{2}{i}\, \GV_{z_0} f_{j-1}. \stepcounter{equation}\tag{\theequation}
\end{align*}
\end{subequations}
This may be shown by substituting Eq.~(\ref{eq:LNseries}) into Eq.~(\ref{eq:LSEreduced2}). That $f_j$ represents the $j$-th order scattering term becomes obvious if Eq.~\ref{eq:LNseries} is rewritten as
\begin{equation}
    \varphi(\bm{r}) = f_0(\bm{r}) + \GV_{z_0}\,f_0(\bm{r})
    + \left(\GV_{z_0}\right)^2 \! f_0(\bm{r}) + \cdots,
\end{equation}
using Eq.~(\ref{eq:fterms}). Eqs.~(\ref{eq:LNseries}) and (\ref{eq:fterms}) are the core connection between LSE and BPM that we will establish in the next section.

\subsection{\label{sec:convBorn} Convergence of the Born series} 
Before discussing the BPM, we briefly take a pause to consider the validity of the Born series. Assuming that solutions of the LSE are continuous, the convergence of the Born series can be shown in a few different ways, e.g. using the Banach-Keissinger theorem \cite{manning1965error}, again given that the operator norm of $\GV_{z_0}$ is less than 1. Otherwise, the convergence of the series cannot be guaranteed and due to the divergent behavior of $\left(\GV_{z_0}\right)^j$ as $n\gg \nb$ and $j \rightarrow \infty$ it would be difficult to obtain the error bound between the series expansion and the true solution of the LSE. Hence, it is important to estimate the dependency of the operator norm on $V$. In other words, we try to estimate conditions on $V$ that make the operator norm of $\GV_{z_0}$ less than 1 in some domain. In numerical computations, we are interested in evaluating $\varphi(\bm{r})$ in a bounded subset $\mathcal{D}$ of $\mathbb{R}^3$, e.g. a box
\begin{equation} \label{eq:LSEdomain}
    \mathcal{D} = \left[-\frac{L_1}{2}, \frac{L_1}{2} \right]
     \times \left[-\frac{L_2}{2}, \frac{L_2}{2} \right] \times
    \left[-\frac{L_3}{2}, \frac{L_3}{2} \right],
\end{equation}
which contains the support of $V$. We now evaluate the operator norm in $\mathcal{D}$.

From the definition of $\GV_{z_0}$, Eq.~(\ref{eq:LSE2DOp2}),
\begin{equation} \label{eq:LSEnormSup}
    \norm{\GV_{\alpha}\varphi}
    \leq
    \norm{\hG} \norm{\varphi} \sup_{\mathcal{D}}\left(V\right)
\end{equation}
where $\norm{\hG}$ is the operator norm of
\begin{equation}
    \hG: \,\, \varphi \rightarrow
    \int_{\mathcal{D}} d\bm{r}^\prime \,\,
    G(\bm{r}-\bm{r}^\prime) \varphi(\bm{r}^\prime).
\end{equation}
It is difficult to get an analytical expression for $\norm{\hG}$, particularly due to the singularity of $G$ at the origin. Instead, \cite{natterer2004error} suggests using a numerical method, which is a crude approximation on the true norm. To achieve a more analytical approach, we first try to remove the singularity using the discussion in \cite{vico2016fast}. It can be easily shown that
\begin{align*}
    &\GV_{z_0}\varphi\\
    &=
    \frac{1}{\psi_0}
    \int_{\mathcal{D}} d\bm{r}^\prime \,\,
    G(\bm{r}-\bm{r}^\prime) 
    \operatorname{rect}
    \left(\frac{\norm{\bm{r}-\bm{r}^\prime}}{2L_M}\right)
    V(\bm{r}^\prime) \psi(\bm{r}^\prime),
    \stepcounter{equation}\tag{\theequation}
\end{align*}
where $L_M$ is the diagonal length of the smallest box containing the support of $V$, e.g. $\sqrt{L_1^2+L_2^2+L_3^2}$. Then $\norm{\hG}$ becomes the norm of a convolution with a new kernel,
\begin{equation}
    \Bar{G}(\bm{r}) = 
    G(\bm{r}) 
    \operatorname{rect}
    \left(\frac{\norm{\bm{r}}}{2L_M}\right),
\end{equation}
whose Fourier transform is entire by virtue of the Paley-Wiener theorem:
\begin{align*}
    \hat{\mathcal{F}} &\Bar{G}(\bm{r})(k)
    =
    \frac{1}{k} \frac{1}{(\nb k_0 -k)(\nb k_0 +k)} 
    \big[ \big. \\
    &\big. \quad
    \ee{i \nb k_0 L_M}(k\cos kL_M - i\nb k_0 \sin kL_M) - k \big].
    \stepcounter{equation}\tag{\theequation}
\end{align*}
Since the Fourier transform is unitary, $\norm{\hG}$ would be bound by the largest Fourier coefficient of $\Bar{G}(\bm{r})$. Under the small wavelength approximation on which the scalar Helmholtz equation is based, $\nb k_0L_M \gg 1$ and subsequently, the absolute value of $\hat{\mathcal{F}} \Bar{G}(\bm{r})(k)$ has two peaks at $k=\nb k_0$ (from surface of momentum conservation) and $k=0$ (from regularization of the singularity), which asymptotically approach $\frac{L_M}{\nb k_0}$ and $\frac{L_M}{2\nb k_0}$, respectively. Therefore,
\begin{equation}
    \norm{\hG} \leq \frac{L_M}{\nb k_0},
\end{equation}
and subsequently,
\begin{equation} \label{eq:LSEnorm}
    \norm{\GV_{z_0}}
    \leq
    \frac{L_M}{\nb k_0}
    \sup_{\mathcal{D}}\left(V\right).
\end{equation}
However, Eq.~(\ref{eq:LSEnorm}) would be too loose an estimate on the operator norm, {\it i.e.} the use of $\underset{\mathcal{D}}{\sup}(V)$ in Eq.~(\ref{eq:LSEnormSup}). Hence we instead suggest using
\begin{equation} \label{eq:LSEnorm2}
    \norm{\GV_{\alpha}}
    \lesssim
    \frac{L_M}{\nb k_0}
    \underset{\mathcal{D}}{\operatorname{mean}}(V)
\end{equation}
as an approximation if the potential $V$ is mostly smooth. Setting $V(\bm{r})=(\nb k_0)^2 \left[ \left(\frac{n(\bm{r})}{\nb}\right)^2 -1 \right]$, Eq.~(\ref{eq:LSEnorm2}) can be rewritten as
\begin{equation} \label{eq:bornnorm}
    \norm{\GV_{\alpha}}
    \lesssim
    L_M \nb k_0 \left[ 
    \left(
    \frac{
    \underset{\mathcal{D}}{\operatorname{mean}}
    \left(n\right)}{\nb}
    \right)^2 -1 
    \right].
\end{equation}
That is, roughly speaking, the validity of the Born series guarantee is inversely proportional to the object scale with respect to the incident wavelength and the square of the refractive index. The estimation of the norm in Eq.~(\ref{eq:bornnorm}) is tighter and simpler than previous reports e.g. \cite{manning1965error,kilgore2017convergence} as the size of optical objects becomes large. A detailed discussion is presented in Appendix \ref{app:BornConv}. The tightness of the bound also helps improve the truncation error estimate expressed as geometric series of the norm, e.g. \cite{manning1965error},
\begin{equation}
    \norm{
    \varphi - 
    \sum_{j=0}^{N}
    \left(\GV_{z_0}\right)^j f_0
    }
    \leq
    \frac{\norm{\GV_{z_0}}^{N+1}}{1-\norm{\GV_{z_0}}}
    \norm{f_0}.
\end{equation}

\section{From Born series to BPM}
\begin{figure}[t]
    \centering
    \includegraphics[width=0.8\linewidth]{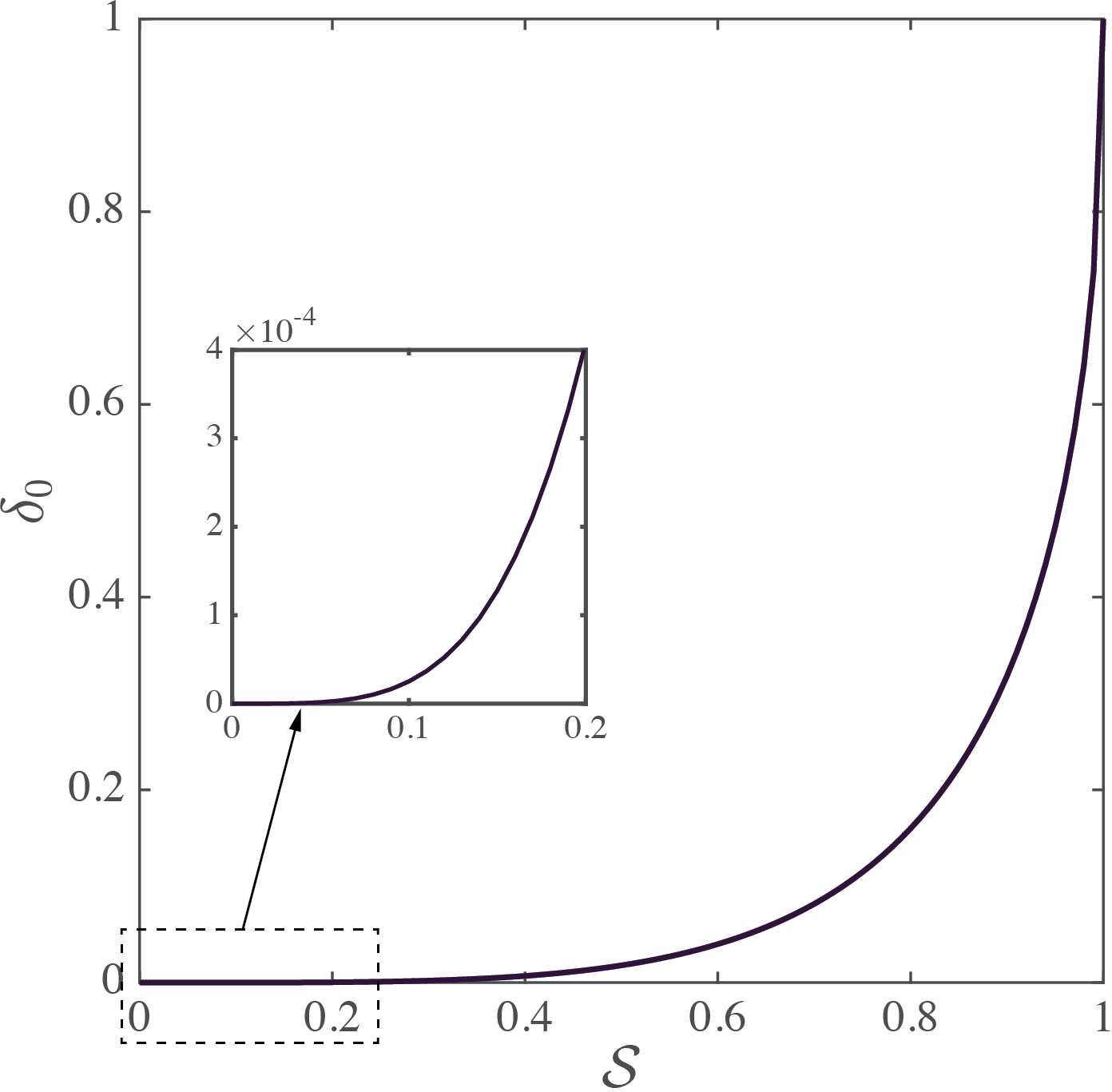}
    \caption{Dependence of $\delta_0$ on $\mathcal{S}$. As $\mathcal{S}$ increases, $\delta_0$ approaches its maximum value, 1. This implies that the Fourier transform of $V$ has significant effects on the validity of the BPM.}
    \label{fig:k0error}
\end{figure}

As discussed in the previous section, Eq.~(\ref{eq:fterms}) plays a key role in connecting LSE and BPM. We begin with analyzing $f_1$, the first term in the Born series, representing a single scattering event:
\begin{align*} \label{eq:born_f1}
    f_1(\bm{r}) 
    &= 
    \int^{z}_{z_0} dz^\prime \,\,
    \hat{\mathcal{F}}_{xy}^\dagger
    \frac{ \ee{i \overline{k}_z (z-z^\prime) } }{k_z}
    \hat{\mathcal{F}}_{xy}
    \Big[ \Big. \\
    & \qquad \Big.
    V(\bm{r}^\prime) 
    \hat{\mathcal{F}}_{xy}^\dagger
    \ee{i \overline{k}_z (z^\prime-z_0) } 
    \hat{\mathcal{F}}_{xy} \left[\varphi(x^\prime,y^\prime,z_0)\right]
    \Big].
    \stepcounter{equation}\tag{\theequation}
\end{align*}
To derive the BPM, it is required that the two operators
\begin{equation} \label{eq:commute_0}
    \hat{\mathcal{F}}_{xy}^\dagger
    \frac{ \ee{i \overline{k}_z (z-z^\prime) } }{k_z}
    \hat{\mathcal{F}}_{xy} 
    \quad \text{and} \quad
    V(\bm{r})\times
\end{equation}
commute. Using the convolution theorem, it can be shown that
\begin{align*}
    \hat{\mathcal{F}}_{xy}^\dagger
    \frac{ \ee{i \overline{k}_z (z-z^\prime) } }{k_z}
    &
    \hat{\mathcal{F}}_{xy} 
    V(\bm{r}^\prime)\\
    &=
    \frac{1}{(2\pi)^2}
    \hat{\mathcal{F}}_{xy}^\dagger
    \frac{ \ee{i \overline{k}_z (z-z^\prime) } }{k_z}
    \left[ \Tilde{V}_{z^\prime} \star \right]
    \hat{\mathcal{F}}_{xy},
    \stepcounter{equation}\tag{\theequation}
\end{align*}
where $\Tilde{V}_{z^\prime} \star$ is a convolution operator:
\begin{equation}
    \Tilde{V}_{z^\prime} \star: \,\,
    \varphi(\bm{k}) 
    \rightarrow
    \int d\bm{k}^\prime
    \hat{\mathcal{F}}_{xy}\left[V(x,y,z)\right](\bm{k}-\bm{k}^\prime)
    \varphi(\bm{k}^\prime).
\end{equation}
Here, we assume that $V$ is band-limited in each of its $xy$-slices. For brevity, we first define the boxcar function in $\mathbb{R}^2$ as
\begin{equation}
    \operatorname{rect}(\bm{x}) = \begin{cases}
    0, & \text{if $\norm{\bm{x}}>\frac{1}{2}$} \\
    1, & \text{otherwise,}
    \end{cases}
\end{equation}
and approximate $\hat{\mathcal{F}}_{xy}\psi$ and $\Tilde{V}_{z^\prime}$ as
\begin{subequations}
\begin{eqnarray}
    \hat{\mathcal{F}}_{xy}\varphi 
    &\approx&
    C_\varphi 
    \operatorname{rect}
    \left(\frac{\bm{k}}{2K_\varphi}\right) \\
    \Tilde{V}_{z^\prime} 
    &\approx&
    C_V 
    \operatorname{rect}
    \left(\frac{\bm{k}}{2K_V}\right),
    \label{eq:potential_rect}
\end{eqnarray}
\end{subequations}
i.e. their support is confined to spheres of size $K_\varphi$ and $K_V$, respectively, while $C_\varphi$ and $C_V$ are upper bounds on the approximate operator amplitudes. It follows that 
\begin{align*} \label{eq:commute_1}
    &\frac{ \ee{i \overline{k}_z (z-z^\prime) } }{k_z}
    \left[ \Tilde{V}_{z^\prime} \star \right]
    \hat{\mathcal{F}}_{xy}\varphi \\
    &\qquad \approx
    C_\varphi C_V
    \frac{ \ee{i \overline{k}_z (z-z^\prime) } }{k_z}
    (\pi K_V^2) 
    \operatorname{rect}\left(\frac{\bm{k}}{2(K_V+K_\varphi)}\right).
    \stepcounter{equation}\tag{\theequation}
\end{align*}
On the other hand, 
\begin{align*} \label{eq:commute_2}
    &\left[ \Tilde{V}_{z^\prime} \star \right]
    \frac{ \ee{i \overline{k}_z (z-z^\prime) } }{k_z} 
    \hat{\mathcal{F}}_{xy}\varphi\\
    &\,\, \approx
    C_\varphi C_V
    \operatorname{rect}\left(\frac{\bm{k}}{2(K_V+K_\varphi)}\right)
    \Bigg[ \ee{-i\nb k_0(z-z^\prime)} \Bigg.\\
    & \qquad \quad \Bigg.
    \int_{B_{K_V}(\bm{k})} d\bm{k}^{\prime}\,\, 
    \frac{ \ee{i(z-z^\prime) 
    \sqrt{(\nb k_0)^2-(k_x^\prime)^2-(k_y^\prime)^2} } }{\sqrt{(\nb k_0)^2-(k_x^\prime)^2-(k_y^\prime)^2}} 
    \,\,
    \Bigg],
    \stepcounter{equation}\tag{\theequation}
\end{align*}
where $B_{K_V}(\bm{k})$ is a ball of radius $K_V$ centered at $\bm{k}$. Comparing Eqs.~(\ref{eq:commute_1}) and (\ref{eq:commute_2}), the two operators in Eq.~(\ref{eq:commute_0}) would commute if
\begin{align*} \label{eq:commute_approx}
    &\pi K_V^2
    \frac{ \ee{i \overline{k}_z (z-z^\prime) } }{k_z}
    \approx
    \ee{-i\nb k_0(z-z^\prime)}\\
    & \qquad \times
    \int_{B_{K_V}(\bm{k})} d\bm{k}^{\prime}\,\, 
    \frac{ \ee{i(z-z^\prime) 
    \sqrt{(\nb k_0)^2-(k_x^\prime)^2-(k_y^\prime)^2} } }{\sqrt{(\nb k_0)^2-(k_x^\prime)^2-(k_y^\prime)^2}},
    \stepcounter{equation}\tag{\theequation}
\end{align*}
{\it i.e.} if the propagator (2D Fourier spectrum of the Green's function) is nearly constant in $B_{K_V}(\bm{k})$ for every $\bm{k}$ in $B_{K_\varphi+K_V}(\bm{0})$. This is consistent with the weak scattering approximation applied separately on each slice of the BPM. To satisfy condition (\ref{eq:commute_approx}), it is sufficient to require that
\begin{equation} \label{eq:bpmvalid}
    \text{$z-z^\prime$ and $K_V$ are small.}
\end{equation}
To further simplify the integrand in Eq.~(\ref{eq:commute_approx}) toward obtaining an estimate of its validity bound, let us assume that $z-z^\prime$ is sufficiently small so that the term $\ee{i \overline{k}_z (z-z^\prime)}$ can be considered locally constant in $B_{K_V}(\bm{k})$ and describe this term as a constant $C_z$. Then, at $\bm{k}=\bm{0}$,
\begin{align*}
    \int_{B_{K_V}(\bm{0})} d\bm{k}^{\prime}\,\,
    &
    \frac{ 
    \ee{i\overline{k}_z(z-z^\prime)} }{
    \sqrt{(\nb k_0)^2-(k_x^\prime)^2-(k_y^\prime)^2}} \\
    & =
    \int_{0}^{2\pi} d\theta \int_{0}^{K_V} r dr 
    \frac{ 
    C_z }{
    \sqrt{k^2-r^2}} \stepcounter{equation}\tag{\theequation} \\ 
    & = 
    2\pi C_z \left(\nb k_0 - \sqrt{(\nb k_0)^2-K_V^2} \right),
\end{align*}
and, subsequently,
\begin{align*} 
    &\Bigg|\pi K_V^2
    \frac{ \ee{i \overline{k}_z (z-z^\prime) } }{k_z}
    \Bigg. \\
    & \qquad - 
    \int_{B_{K_V}(\bm{0})} d\bm{k}^{\prime}\,\,
    \frac{ \ee{i\overline{k}_z(z-z^\prime) } }{\sqrt{(\nb k_0)^2-(k_x^\prime)^2-(k_y^\prime)^2}}
    \Bigg| \\
    &\approx 
    \pi C_z \nb k_0
    \left(2 - 2\sqrt{1-\mathcal{S}^2} - \mathcal{S}^2  \right),
    \stepcounter{equation}\tag{\theequation}
    \label{eq:hereisdelta}
\end{align*}
where $\mathcal{S}$ is the dimensionless parameter
\begin{equation}
    \mathcal{S} \equiv \frac{K_V}{\nb k_0}.
\end{equation}
We shall refer to the last term in Eq.~(\ref{eq:hereisdelta}) as
\begin{equation} \label{eq:small_S}
    \delta_0 = 
    2 - 2\sqrt{1-\mathcal{S}^2} - \mathcal{S}^2 
    \approx 
    \frac{{\cal S}^4}{2}.
\end{equation}
The behavior of $\delta_0$ {\it vs.} ${\cal S}$ is shown further down in Fig.~\ref{fig:k0error} as part of a longer discussion on the BPM's validity. The approximation applies for ${\cal S}\ll 1$.

From Eqs.~(\ref{eq:commute_1}) and (\ref{eq:commute_2}), Eq.~(\ref{eq:hereisdelta}) corresponds to the error of the commutation at $\bm{k}=\bm{0}$ (more precisely, the error normalized by $C_\varphi$ and $C_V$ that are average amplitudes of $\varphi$ and $V$ in the Fourier space). When $\bm{k}\neq \bm{0}$ it is not straightforward to derive an analytical expression for the error, but we can anticipate that it would be proportional to $\norm{\bm{k}}$. This is because $ 1/\sqrt{(\nb k_0)^2-(k_x^\prime)^2-(k_y^\prime)^2}$ in Eq.~(\ref{eq:commute_2}) changes rapidly as the domain of integral, $B_{K_V}(\bm{k})$, moves away from the origin in the Fourier space.
Hence, 
\begin{align*} \label{eq:commute_error_2}
    &\left|
    \frac{ \ee{i \overline{k}_z (z-z^\prime) } }{k_z}
    \left[ \Tilde{V}_{z^\prime} \star \right]
    \hat{\mathcal{F}}_{xy}\varphi
    -
    \left[ \Tilde{V}_{z^\prime} \star \right] 
    \frac{ \ee{i \overline{k}_z (z-z^\prime) } }{k_z} 
    \hat{\mathcal{F}}_{xy}\varphi
    \right|\\
    &\approx
    \underbrace{\pi C_\varphi C_z C_V \nb k_0 \delta_0}_{\varepsilon_0}
    + \varepsilon\left(K_V,K_\varphi\right), \stepcounter{equation}\tag{\theequation}
\end{align*}
where  $\varepsilon$ represents the additional error originating from $\bm{k}\neq \bm{0}$ regions, which depends on the effective support of both $V$ and $\varphi$ in the Fourier space and increases more rapidly than $\varepsilon_0$. 


From now on, assume that Eq.~(\ref{eq:bpmvalid}) is satisfied in our system. Then, Eq.~(\ref{eq:born_f1}) becomes
\begin{align*}
    &f_1(\bm{r}) = \\
    &= 
    \int^{z}_{z_0} dz^\prime \,\,
    V(x,y,z^\prime)
    \hat{\mathcal{F}}_{xy}^\dagger
    \frac{ \ee{i \overline{k}_z (z-z_0) } }{k_z}
    \hat{\mathcal{F}}_{xy} 
    \left[\varphi(x,y,z_0)\right] \\
    &= 
    \left\{
    \int^{z}_{z_0} dz^\prime \,\,
    V(x,y,z^\prime)
    \right\}
    \hat{\mathcal{F}}_{xy}^\dagger
    \frac{ \ee{i \overline{k}_z (z-z_0) } }{k_z}
    \hat{\mathcal{F}}_{xy} 
    \left[\varphi(x,y,z_0)\right].
    \stepcounter{equation}\tag{\theequation}
\end{align*}
Subsequently, evaluating $f_2$ yields
\begin{align*}
    &f_2(\bm{r}) = \\
    &= 
    \int^{z}_{z_0} dz^\prime \,\,
    \hat{\mathcal{F}}_{xy}^\dagger
    \frac{ \ee{i \overline{k}_z (z-z^\prime) } }{k_z}
    \hat{\mathcal{F}}_{xy} 
    \Bigg[ \Bigg.
    V(\bm{r}^\prime)
    \left\{
    \int^{z^\prime}_{z_0} dz^{\prime\prime} \,\,
    V(\bm{r}^{\prime\prime})
    \right\} \\
    & \qquad \qquad \times \Bigg.
    \hat{\mathcal{F}}_{xy}^\dagger
    \frac{ \ee{i k_z (z^\prime-z_0) } }{k_z}
    \hat{\mathcal{F}}_{xy} 
    \left[\varphi(x,y,z_0)\right]
    \Bigg]\\
    &= 
    \left\{
    \int^{z}_{z_0} dz^\prime \,\,
    V(x,y,z^\prime)
    \int^{z^\prime}_{z_0} dz^{\prime\prime} \,\,
    V(x,y,z^{\prime\prime})
    \right\} \\
    & \qquad \qquad \times 
    \hat{\mathcal{F}}_{xy}^\dagger
    \frac{ \ee{i \overline{k}_z (z-z_0) } }{k_z^2}
    \hat{\mathcal{F}}_{xy} 
    \left[\varphi(x,y,z_0)\right] \\
    &= 
    \frac{1}{2!}
    \left\{
    \int^{z}_{z_0} dz^\prime \,\,
    V(x,y,z^\prime)
    \right\}^2 \\
    & \qquad \qquad \times 
    \hat{\mathcal{F}}_{xy}^\dagger
    \frac{ \ee{i \overline{k}_z (z-z_0) } }{k_z^2}
    \hat{\mathcal{F}}_{xy} 
    \left[\varphi(x,y,z_0)\right],
    \stepcounter{equation}\tag{\theequation}
\end{align*}
where the last equality is derived using integration-by-parts \cite{ishizuka1977new}. Repeating the same procedure, we can deduce
\begin{align*}
    f_j(\bm{r}) &= 
    \frac{1}{j!}
    \left\{
    \int^{z}_{z_0} dz^\prime \,\,
    V(x,y,z^\prime)
    \right\}^j\\
    & \qquad \times
    \hat{\mathcal{F}}_{xy}^\dagger
    \frac{ \ee{i \overline{k}_z (z-z_0) } }{k_z^j}
    \hat{\mathcal{F}}_{xy} 
    \left[\varphi(x,y,z_0)\right].
    \stepcounter{equation}\tag{\theequation}
\end{align*}
From the analysis on the commutation error, BPM requires $K_\varphi$ and $K_V$ to be small. Hence, $|k_x|, |k_y| \ll \nb k_0$ and $k_z \approx \nb k_0$. Subsequently,
\begin{align*} \label{eq:LNseries_bpm}
    f_j(\bm{r}) &\approx 
    \frac{1}{j!}
    \left\{
    \int^{z}_{z_0} dz^\prime \,\,
    V(x,y,z^\prime)
    \right\}^j\\
    & \qquad \times
    \hat{\mathcal{F}}_{xy}^\dagger
    \frac{ \ee{i \overline{k}_z (z-z_0) } }{(\nb k_0)^j}
    \hat{\mathcal{F}}_{xy} 
    \left[\varphi(x,y,z_0)\right].
    \stepcounter{equation}\tag{\theequation}
\end{align*}
Inserting Eq.~(\ref{eq:LNseries_bpm}) to Eq.~(\ref{eq:LNseries}) gives
\begin{align*} \label{eq:bpm}
    \varphi(\bm{r}) &= 
    \exp\left(
    \frac{i}{2\nb k_0} \left\{
    \int^{z}_{z_0} dz^\prime \,\,
    V(x,y,z^\prime)
    \right\}
    \right) \\
    & \qquad \times
    \hat{\mathcal{F}}_{xy}^\dagger
    \ee{i \overline{k}_z (z-z_0) } 
    \hat{\mathcal{F}}_{xy} 
    \left[\varphi(x,y,z_0)\right] \\
    &= 
    \exp\left(
    \frac{i \nb k_0 }{2} \left\{
    \int^{z}_{z_0} dz^\prime \,\,
    \left[ \left(\frac{n(x,y,z^\prime)}{\nb}\right)^2 -1 \right]
    \right\}
    \right)\\
    & \qquad \times
    \hat{\mathcal{F}}_{xy}^\dagger
    \ee{i \overline{k}_z (z-z_0) }
    \hat{\mathcal{F}}_{xy} 
    \left[\varphi(x,y,z_0)\right] \\
    &\approx 
    \exp\left(
    \frac{i \nb k_0 }{\xi} (z-z_0)
    \left[ \left(\frac{n(x,y,z)}{\nb}\right)^{\xi} -1 \right]
    \right) \\
    & \qquad \times
    \hat{\mathcal{F}}_{xy}^\dagger
    \ee{i \overline{k}_z (z-z_0) }
    \hat{\mathcal{F}}_{xy} 
    \left[\varphi(x,y,z_0)\right],
    \stepcounter{equation}\tag{\theequation}
\end{align*}
where $\xi=2$. Comparing Eqs.~(\ref{eq:fterms}) and (\ref{eq:LNseries_bpm}), it is implied that the $j$-th order scattering term in Born series corresponds to the $j$-th order polynomial in the Taylor expansion of the exponential modulation in the BPM.

\subsection{Difference between Born series and BPM \label{sec:born_bpm}}
Though Born series and BPM both originate from the LSE and their mathematical structures are closely related, BPM imposes different assumptions on the scattering process. First, due to Eq.~(\ref{eq:bpmvalid}), it is required that $|z-z_0|$ be small. Hence, previous studies on BPM suggest slicing a thick $V$ along the optical axis and applying BPM on each slice consecutively. However, this violates our assumption that $z$ is outside of the support of $V$, as in Fig.~\ref{fig:basic_geom}. In other words, at each $j^{\text{th}}$ slice inside $V$, BPM has a numerical discrepancy
\begin{equation} \label{eq:bpm_backscatter}
    \frac{i}{2} \int^{z}_{z_0} dz^\prime \,\,
    \hat{\mathcal{F}}_{xy}^\dagger
    \frac{ \ee{i \overline{k}_z (z-z^\prime) } }{k_z}
    \hat{\mathcal{F}}_{xy} 
    \bigg[
    V(\bm{r}^\prime)  
    \big[\varphi_{j-1}-\varphi\big](\bm{r}^\prime)
    \bigg],
\end{equation}
where $\varphi_{j-1}$ is a field at the $(j-1)^{\text{th}}$ slice in BPM and $\varphi$ is that of LSE. The difference $\varphi_{j-1}-\varphi$ would approximately amount to backscattered fields from $V(x,y,z)$ where $z \geq z_j$ and $z_j$ is the $z$-coordinate of the $j^{\text{th}}$ slice.

Despite Eq.~(\ref{eq:bpm}) suggesting a close connection between Born series and BPM, they do exhibit different numerical convergence. Specifically, BPM is known to be numerically stable with high $V$, compared to the Born series. We may be able to speculate that such behavior can be attributed to the following conditions. First, in BPM, it is assumed that $K_\varphi$ and $K_V$ are small, which makes $1/k_z$ as small as possible in the expansion. In other words, all Fourier coefficients that are multiplied with large $1/k_z$ are effectively ignored, and that  promotes convergence. Second, as in Eq.~(\ref{eq:bpm_backscatter}), BPM does not consider backscattered fields. This would decrease the norm of the LSE operator. We present numerical experiments on comparing the convergence behavior of Born series and BPM in Sec.~\ref{sec:num_disc}.

\subsection{On the appearance of a different value of $\xi$ \\ in BPM's wave modulation term \label{sec:bpm_xi_value}}
According to Eq.~(\ref{eq:bpm}), BPM consists of two operations. First, an incident field is propagated with small distance $z-z_0$. Subsequently, the field undergoes a phase modulation. The modulation is proportional to $(n/\nb)^\xi/\xi$ where $\xi=2$. This resembles BPM in previous studies except they suggest $\xi=1$ \cite{paganin2006coherent,feit1988beam}. 

The difference in the assumed values of $\xi$ originates from the respective assumptions. To track the differences, let us again start with the Helmholtz equation Eq.~(\ref{eq:helmholtz}), rewritten here for convenience as 
\begin{equation}
    \left[
    \frac{\partial^2}{\partial^2 z} +
    \nabla_{xy}^2 + k_0^2 n^2
    \right]\psi = 0,
\end{equation}
where $\nabla_{xy}$ refers to the gradient in the lateral dimensions. Setting $\hat{P}^2=\frac{\partial}{\partial z}$ and $\hat{Q}^2=\nabla_{xy}^2 + k_0^2 n^2$, the equation can be further simplified as 
\begin{equation}
    \left[ 
    (\hat{P}+i\hat{Q})(\hat{P}-i\hat{Q}) 
    + i\left\langle P,Q \right\rangle
    \right] \psi = 0,
\end{equation}
where $\langle , \rangle$ is the commutator. If the variation of $n$ along the optical axis is negligible, then $\left\langle P,Q \right\rangle \rightarrow 0$ \cite{feit1988beam}, which requires
\begin{equation} \label{eq:P_iQ}
    \left[ \hat{P}-i\hat{Q} \right]\psi = 0.
\end{equation}
In fact, we have another set of solutions from $\left[ \hat{P}+i\hat{Q} \right]\psi = 0$, but this represents fields propagating backwards \cite{teague1983deterministic}. Consequently, from Eq.~(\ref{eq:P_iQ}), $\psi$ can be expressed as
\begin{equation}
    \psi(x,y,z) = \exp 
    \left[ 
    i(z-z_0)
    \left(\nabla_{xy}^2 + k_0^2n^2 \right)^{1/2} 
    \right] \psi(x,y,z_0).
\end{equation}
To derive the BPM, it is required to separate $\nabla_{xy}^2$ from $n^2$ in the square root. A straightforward way to separate them is to use the Taylor expansion:
\begin{align*} \label{eq:par_helm_1}
    \left(\nabla_{xy}^2 + k_0^2n^2 \right)^{1/2} 
    &=
    k_0 \left(
    1 + \frac{1}{k_0^2}\nabla_{xy}^2 + (n^2-1) 
    \right)^{1/2} \\
    &\approx
    k_0 + \frac{1}{2k_0} \nabla_{xy}^2 + \frac{k_0}{2}(n^2-1).
    \stepcounter{equation}\tag{\theequation}
\end{align*}
Eq.~(\ref{eq:par_helm_1}) would be satisfied if $\norm{\frac{1}{k_0^2}\nabla_{xy}^2 + (n^2-1) }$ is small, i.e. both the refraction angle and the lateral variation of $n$ are small \cite{thomson1983wide}. Eq.~(\ref{eq:par_helm_1}) corresponds to the phase modulation with $\xi=2$, which uses the same assumptions on fields leading to the derivation of Eq.~(\ref{eq:bpm}). On the other hand, \cite{feit1978light,feit1988beam} suggest that
\begin{equation} \label{eq:par_helm_2}
    \left(\nabla_{xy}^2 + k_0^2n^2 \right)^{1/2}
    \approx
    (\nabla_{xy}^2 + k_0^2)^{1/2} + k_0(n-1),
\end{equation}
which can be justified if the lateral variation of $n$ is small. This corresponds to the phase modulation with $\xi =1$. 

Summarizing, Eqs.~(\ref{eq:par_helm_1}) for $\xi=2$ and (\ref{eq:par_helm_2}) for $\xi=1$ require different assumptions. The former requires both $\nabla_{xy}^2\psi$ and $\nabla_{xy}^2n$ to be small; whereas the latter does not need the small refraction angle condition. However, the small lateral variation of $n$ indirectly implies that the refraction angle of $\psi$ in the potentials also needs to be small. Hence, it is expected that the $\xi=1$ modulation would not result in significant difference over the $\xi=2$ modulation, especially when $\mathcal{S}$ is small. This was confirmed empirically by our numerical observations. Explicitly, we depict the effect of $\xi$ on spherical potentials in Appendix~\ref{sec:xi}.

\subsection{Validity of the BPM}
Eqs.~(\ref{eq:bpmvalid}) and (\ref{eq:commute_error_2}) imply that the BPM approaches the LSE as $K_V$, the upper bound of diffraction away from the optical axis, becomes smaller. Hence, the difference between BPM and LSE would also depend on $K_V$ and $\mathcal{S}$. Since, again, the exact evaluation of such difference can be difficult, here we devise some simplifying approximations that also lend some insight to the problem. From Eq.~(\ref{eq:potential_rect}),
\begin{align*} \label{eq:V_from_rect}
    V_z(\bm{x}) 
    &\approx C_V K_V^2 \operatorname{sinc}\left(\rule[-1ex]{0cm}{2ex} 2K_V\!\norm{\bm{x}}\right) \\
    &\approx (k_0\nb)^2\left(\frac{n_z(\bm{x})}{\nb}\right)^2
    \stepcounter{equation}\tag{\theequation}
\end{align*}
where the subscript $z$ is used to represent a $z$-slice. In other words, $V$ is a function whose amplitude is $(k_0n_z)^2$ and effective support is $K_V^{-1}$. Assuming that the gradient of $n_z$ in the $xy$ plane is negligible, we may derive 
\begin{equation}
    \varepsilon_0
    \approx
    C_\varphi C_z \mathcal{S}^{-2} 
    \left(\frac{n_z}{\nb}\right)^2
    (\nb k_0) \delta_0.
\end{equation}
This is the commutation error at $\bm{k}=\bm{0}$ in Eq.~(\ref{eq:commute_error_2}). If $\mathcal{S}$ is sufficiently small, Eq.~(\ref{eq:small_S}) gives
\begin{equation}
    \varepsilon_0 \approx
    C_\varphi C_z 
    \left(\frac{n_z}{\nb}\right)^2 (\nb k_0)\: \mathcal{S}^{2}.
\end{equation}
Neglecting the diffraction effect between $z$ and $z_0$, the commutation error in the first order scattering term, Eq.~(\ref{eq:born_f1}),  becomes
\begin{align*} \label{eq:comm_error_dz}
    \varepsilon_{z,z_0} &= 
    \int^{z}_{z_0} dz^\prime \,\,
    \hat{\mathcal{F}}_{xy}^\dagger
    \left[ 
    C_\varphi C_z 
    \left(\frac{n_{z^\prime}}{\nb}\right)^2 
    (\nb k_0) \mathcal{S}^{2}
    +\varepsilon
    \right] \\
    & \approx
    (z-z_0)
    \hat{\mathcal{F}}_{xy}^\dagger
    \left[ 
    C_\varphi C_z 
    \left(\frac{n_{z_0}}{\nb}\right)^2 
    (\nb k_0) \mathcal{S}^{2}
    +\varepsilon
    \right],
    \stepcounter{equation}\tag{\theequation}
\end{align*}
where the subscripts in $\varepsilon_{z,z_0}$ are used to emphasize that now we consider the total commutation error from a potential slice. If we approximate $\varepsilon$ as a function whose amplitude is $\varepsilon_0$ and effective support is mostly governed by $\varphi$, then Eq.~(\ref{eq:comm_error_dz}) finally becomes
\begin{equation}
    \varepsilon_{z,z_0} \approx
    C (z-z_0) 
    \left(\frac{n_{z_0}}{\nb}\right)^2 (\nb k_0)
    \:\mathcal{S}^{2},
\end{equation}
where $C$ is a dimensionless number that is almost independent of the system configuration. In addition, since we require $\ee{i \overline{k}_z (z-z_0)}$ to be nearly constant in the derivation of BPM, $\nb k_0(z-z_0) $ can be regarded as another dimensionless number that is independent of the system configuration. Subsequently, we can further simplify $\varepsilon_{z,z_0}$ as
\begin{equation}
    \varepsilon_{z,z_0} \approx
    C \left(\frac{n_{z_0}}{\nb}\right)^2
    \mathcal{S}^{2}.
\end{equation}
Using $\varepsilon_{z,z_0}$, the total commutation error, $\varepsilon_t$, in the first order scattering term from an entire potential can be expressed. Let us denote as $z_1, \cdots, z_N$ th locations of the $z$-slices along the optical axis. Then
\begin{align*} \label{eq:total_error}
    \varepsilon_t 
    & = \sum_{m=1}^{N} \varepsilon_{z_m,z_{m-1}} \\
    & = C \sum_{m=1}^{N} \left(\frac{n_{z_{m-1}}}{\nb}\right)^2
        \mathcal{S}^{2}(z_{m-1}) \\
    & \approx 
    C (\nb k_0) \int^{R_z/2}_{R_z/2} dz \,\,
    \left(\frac{n_{z}}{\nb}\right)^2
    \mathcal{S}^{2}(z) 
    \stepcounter{equation}\tag{\theequation}
\end{align*}
where the $z$ dependency of $\mathcal{S}$ is due to $K_V$ in $\mathcal{S}$, and that is approximately reciprocal to the size of the potential in the $xy$ plane; whereas $R_z$ is the size of the potential along the optical axis. 

Eq.~(\ref{eq:total_error}) implies that the error of BPM increases as the thickness of the potential increases and the lateral size of the potential decreases, which agrees with previous studies on optical scattering. What is important is that the effect of the lateral size is larger than that of the thickness. To be more specific, we can consider a case of Mie scattering where an incident planewave is scattered by a spherical potential of radius $R_z$ with constant refractive index $n$. Then
\begin{equation}
    K_V(z) \sim \frac{1}{\sqrt{R_z^2-z^2}}, 
    \quad z \in \left[-\frac{R_z}{2}, \frac{R_z}{2} \right], 
\end{equation}
which gives
\begin{equation}
    \varepsilon_t
    \approx C \left(\frac{n}{\nb}\right)^2
    \frac{1}{\nb k_0R_z}
    \operatorname{ln}3.
\end{equation}
In other words, as the sphere becomes large with respect to the incident wavelength, the error decreases though the thickness of the potential grows. This is because the average error at each potential slice decreases more rapidly.

Overall, Eq.~(\ref{eq:total_error}) entails that BPM approximates the LSE if the magnitude of the refractive index $n$ and the dimensionless parameter $\mathcal{S}$ are both small enough. Qualitatively, small $\mathcal{S}$ implies that the variation of $n$ along the lateral direction should be small in the scale of the wavelength. In addition, Eq.~(\ref{eq:bpm}) suggests that the variation of $n$ should also be small along the optical axis. These ideas agree with previous studies \cite{feit1978light,feit1988beam}. Due to the complex behavior of $\varepsilon$ and the accumulation of commutation error in high order scattering terms in Eq.~(\ref{eq:LNseries_bpm}), the actual dependency of the difference between BPM and LSE may deviate from $\varepsilon_t$. Nevertheless, it can serve as a useful lower bound for the accuracy of the BPM.

\section{Numerical discussion} \label{sec:num_disc}

\begin{figure}[t]
\centering
\includegraphics[width=\linewidth]{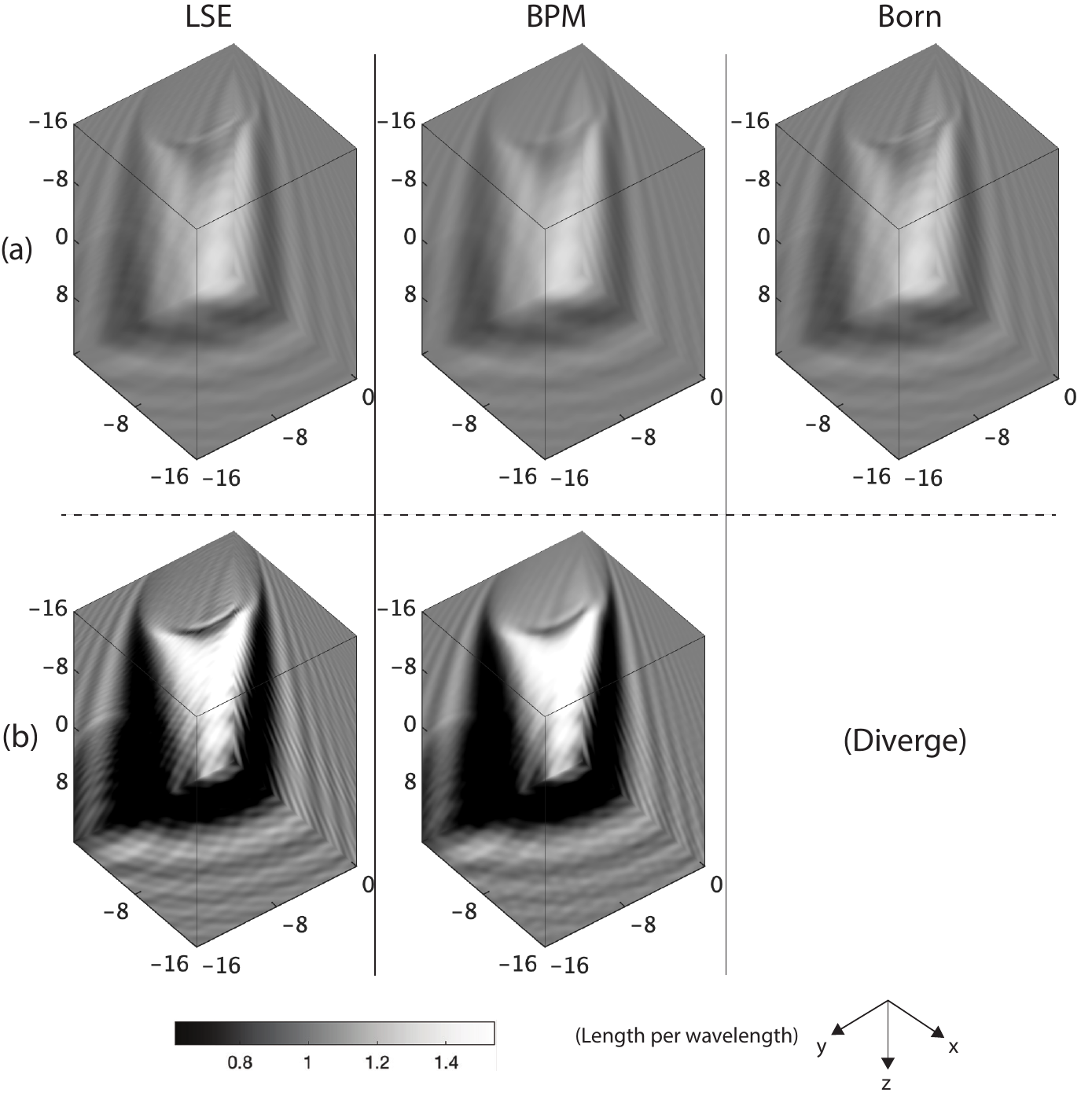}
\caption{\label{fig:born_conv} Comparison of scattered fields from LSE, BPM, and Born series. Two different dielectric spheres are considered where we only change $n$ to adjust the estimated norm of the LSE operator in Eq.~(\ref{eq:bornnorm}). (a) The norm is 0.9. (b) The norm is 15. }
\end{figure}

In this section, we try to numerically validate our discussions on LSE, Born series, and BPM. Before proceeding further, we first demonstrate that LSE well approximates the finite-difference time-domain (FDTD) solutions in Appendix~\ref{app:FDTD}.

In Sec.~\ref{sec:born_bpm}, we discuss the stronger convergence behavior of BPM compared to Born series. Mainly, this is because BPM neglects high $1/k_z$ portions in the field propagator, though both methods originate from the same polynomial series of $f_j$. Fig.~\ref{fig:born_conv} shows how scattered field estimations depend on the magnitude of $n$. As $n$ increases, the upper bound of the operator norm of the LSE operator in Eq.~(\ref{eq:bornnorm}) becomes high, which indicates the divergence of Born series. On the other hand, BPM does not exhibit such divergence.

\begin{figure*}
\centering
\includegraphics[width=\textwidth]{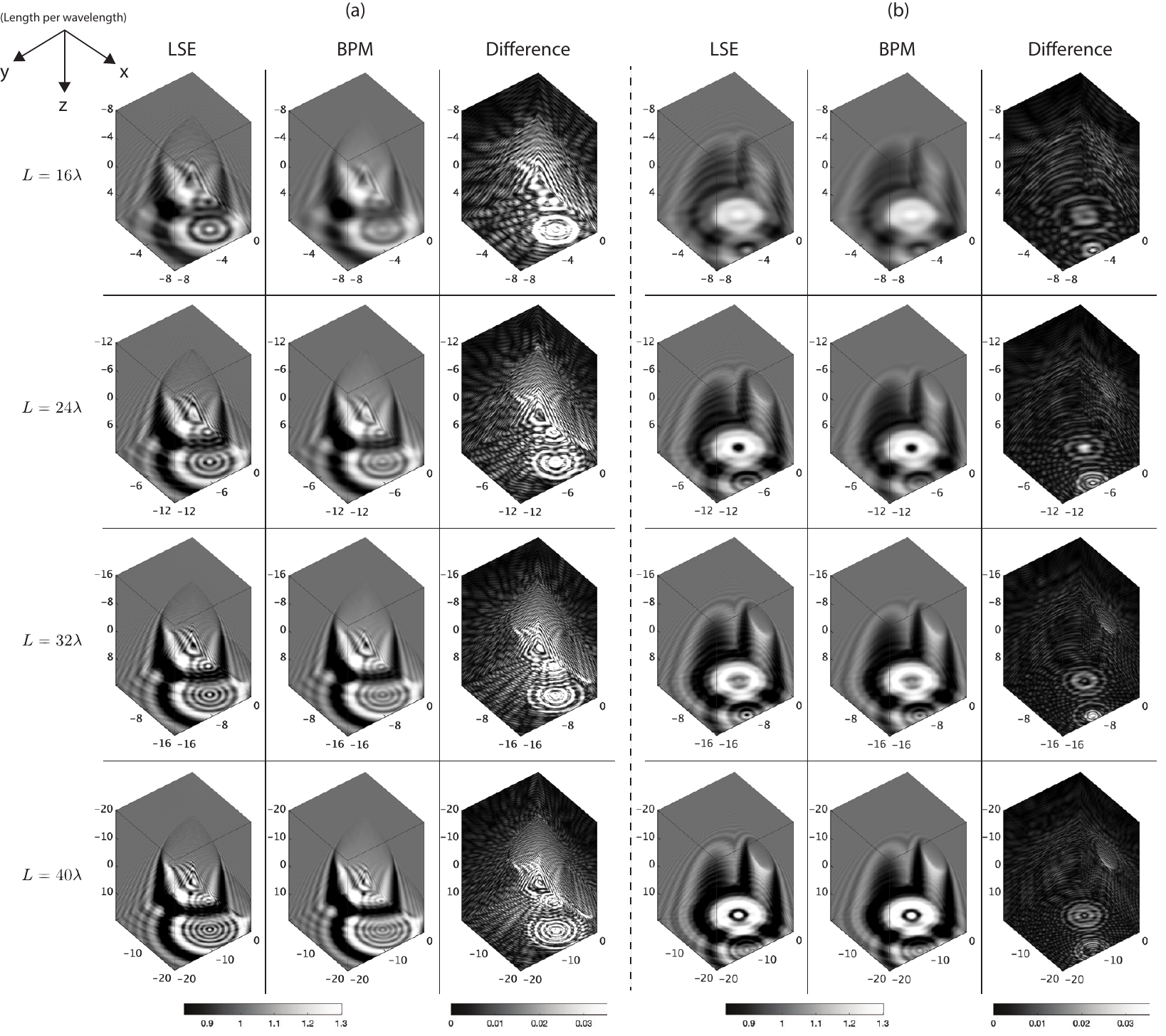}
\caption{\label{fig:index_1_02} Scattered fields estimated from LSE and BPM when the size $L$ of a cubic computational box changes. We consider two distinct potentials, marked as (a) and (b), both consisting of dielectric spheres. The mean refractive index is $1.02$. The difference refers to the elementwise absolute error divided by the maximum field amplitude.}
\end{figure*}

\begin{figure*}
\centering
\includegraphics[width=\textwidth]{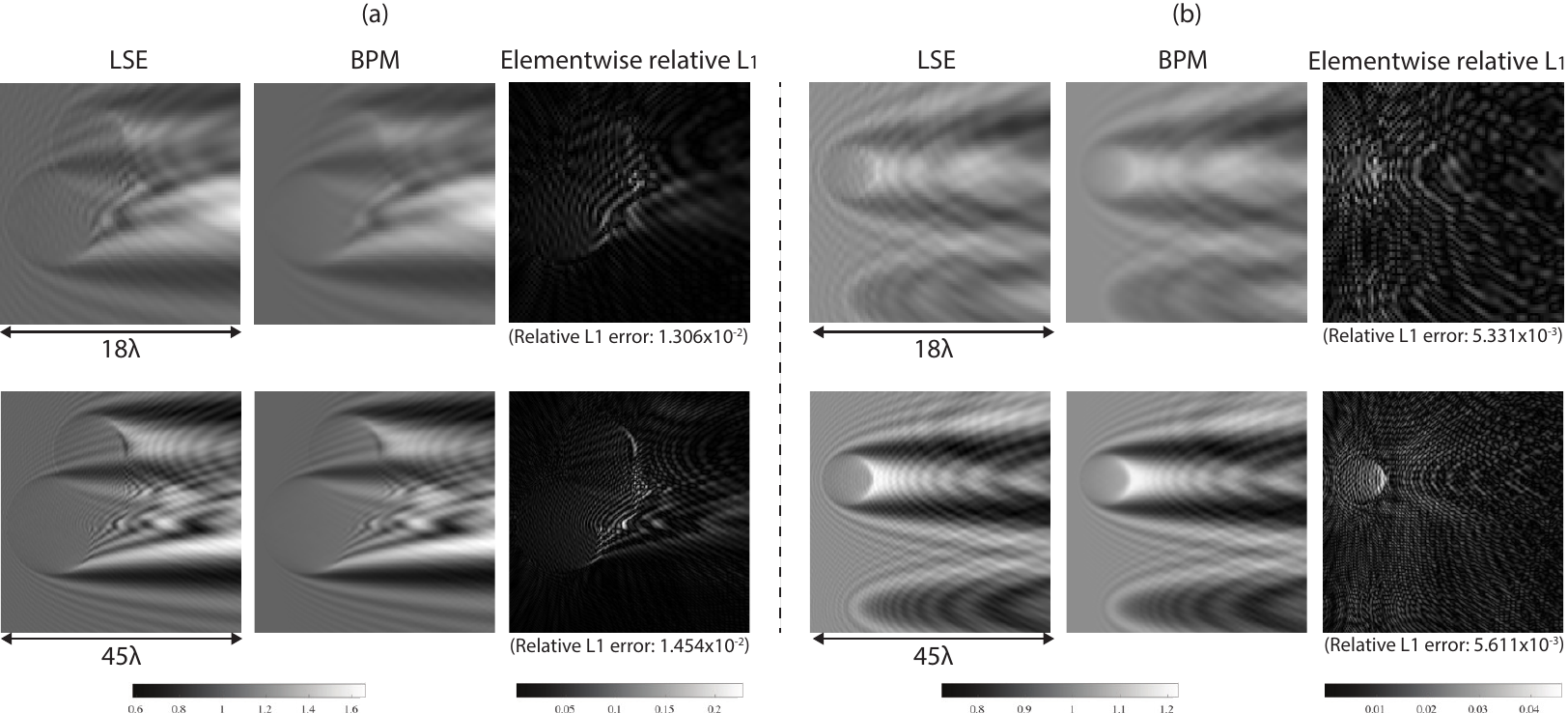}
\caption{\label{fig:index_1_02_xz} $xz$-view of scattered fields estimated from LSE and BPM for the objects as in Fig.~\ref{fig:index_1_02}, marked as (a) and (b).}
\end{figure*}

\begin{figure*}
\centering
\includegraphics[width=\textwidth]{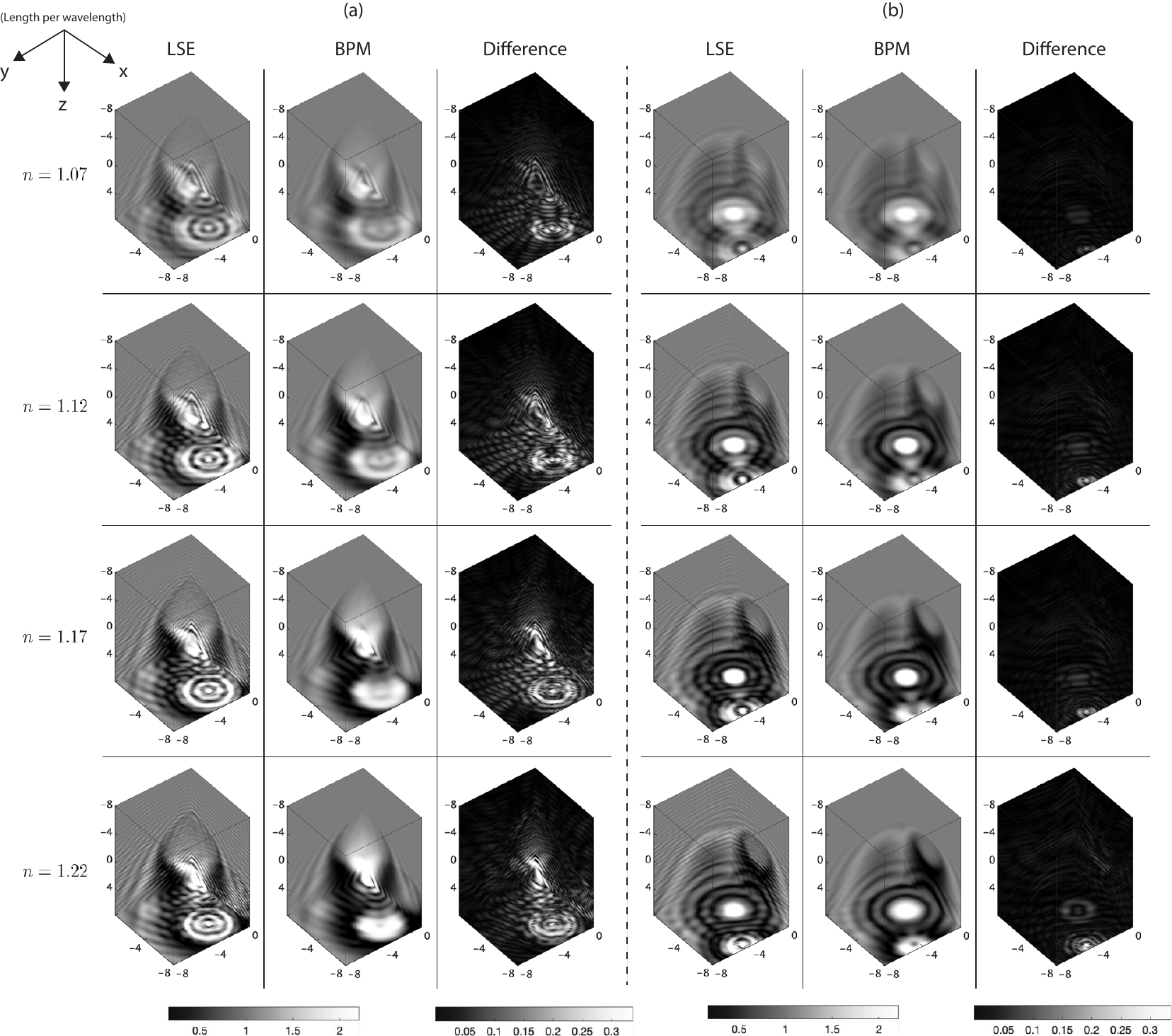}
\caption{\label{fig:scale_4} Scattered fields estimated from LSE and BPM when the mean refractive index $n$ of spherical potentials changes. We consider potentials which consist of spheres. The size of a cubic computational box is $16\lambda$. We show two different objects, which are marked with (a) and (b). Difference refers to the elementwise absolute error divided by the maximum field amplitude.}
\end{figure*}

We further investigate the difference between LSE and BPM. Qualitatively speaking, it is controlled by the dimensionless parameter $\mathcal{S}$, which tells that large size and small refractive index induce small difference. In Fig.~\ref{fig:index_1_02}, we can see that complex interference patterns near small objects are not well estimated in BPM. We also present quantitative comparison between them in Table~\ref{tab:1_02_scale} by measuring the structural similarity index (SSIM) \cite{wang2004image}, the peak signal-to-noise ratio (PSNR) \cite{hore2010image} and the relative $L_1$ error (also referred to as MAE, mean absolute error.) The quantitative metrics follow the same trend as the qualitative analysis, except the $L_1$ error in amplitude. This can be attributed to high frequency oscillations along the optical axis when $\psi_0$ is scattered by relatively large objects. For example, in Fig.~\ref{fig:index_1_02_xz}, we again see the good agreement between LSE and BPM as the size of potentials increases. At the same time, fine stripes of high relative $L_1$ errors appear, which originates from oscillatory patterns in amplitudes along the optical axis. Such patterns are numerically subtle to estimate accurately. On the other hand, Fig.~\ref{fig:scale_4} and Table~\ref{tab:nidx} demonstrates strong reciprocity between the magnitude of the refractive index and the error between LSE and BPM, which agrees with our theoretical analysis. As additional information, we present the size dependency of the error between LSE and BPM with relatively high mean refractive indices in Appendix \ref{sec:LSE_BPM_addition}.

\begin{table}[t]
\caption{\label{tab:1_02_scale} Image quality metrics on fields from LSE and BPM when the size $L$ of a cubic computational box changes. We consider 15 different potentials which consist of dielectric spheres. The mean refractive index is $1.02$. The phase is unwrapped along the optical axis. The full width at half maximum of the Gaussian window in SSIM is $\lambda/2$.}
\begin{ruledtabular}
\begin{tabular}{cccc}
&
\textrm{SSIM}&
\textrm{PSNR}&
\textrm{Relative $L_1$}\\
\colrule
$L=16\lambda$, amplitude & $0.948$ & $37.996$ & $6.683\times 10^{-3}$\\
$L=24\lambda$, amplitude & $0.965$ & $40.147$ & $6.909\times 10^{-3}$\\
$L=32\lambda$, amplitude & $0.974$ & $41.732$ & $7.127\times 10^{-3}$\\
$L=40\lambda$, amplitude & $0.977$ & $42.616$ & $7.400\times 10^{-3}$\\
$L=16\lambda$, phase & $0.991$ & $38.067$ & $4.101\times 10^{-2}$\\
$L=24\lambda$, phase & $0.995$ & $41.617$ & $2.698\times 10^{-2}$\\
$L=32\lambda$, phase & $0.997$ & $44.126$ & $2.010\times 10^{-2}$\\
$L=40\lambda$, phase & $0.998$ & $46.067$ & $1.602\times 10^{-2}$\\
\end{tabular}
\end{ruledtabular}
\end{table}

\begin{table}[t]
\caption{\label{tab:nidx} Image quality metrics on fields from LSE and BPM when the mean refractive index $n$ of spherical potentials changes. We consider 15 different potentials which consist of dielectric spheres. The size of the cubic computational box is $16\lambda$. The phase is unwrapped along the optical axis. The full width at half maximum of the Gaussian window in SSIM is $\lambda/2$.}
\begin{ruledtabular}
\begin{tabular}{cccc}
&
\textrm{SSIM}&
\textrm{PSNR}&
\textrm{Relative $L_1$}\\
\colrule
$n=1.07$, amplitude & $0.931$ & $36.722$ & $2.790\times 10^{-2}$\\
$n=1.12$, amplitude & $0.888$ & $34.429$ & $6.291\times 10^{-2}$\\
$n=1.17$, amplitude & $0.838$ & $32.394$ & $9.715\times 10^{-2}$\\
$n=1.22$, amplitude & $0.812$ & $31.137$ & $12.076\times 10^{-2}$\\
$n=1.07$, phase & $0.990$ & $39.126$ & $4.114\times 10^{-2}$\\
$n=1.12$, phase & $0.971$ & $36.198$ & $4.339\times 10^{-2}$\\
$n=1.17$, phase & $0.933$ & $31.826$ & $5.272\times 10^{-2}$\\
$n=1.22$, phase & $0.910$ & $29.105$ & $6.042\times 10^{-2}$
\end{tabular}
\end{ruledtabular}
\end{table}

\section{Conclusions}
In this work, we discuss analytical relationships between three methods for estimating optical scattering: LSE, BPM, and Born series. It is shown that BPM and Born series both can originate from the series expansion of LSE. However, they exhibit different convergence behavior. Analyzing this behavior, we suggest a simple and dimensionless condition to guarantee the convergence of Born series that is tighter than previous studies. Furthermore, assumptions behind BPM that field propagation and modulation from optical potentials commute can effectively reduce the operator norm of the LSE operator, leading to a stronger convergence than Born series. The errors resulting from such commutation assumption can be estimated by a dimensionless parameter $\mathcal{S}$. Subsequently, we conduct numerical experiments, which corroborate the feasibility of our theoretical analysis. We limited our analysis to scattering from the Helmholtz model; we expect that the discussions are applicable to other scattering models, relevant methods and experimental conditions.

\begin{acknowledgments}
This research was funded by the Intelligence Advanced Research Projects Activity (IARPA) as part of the Rapid Analysis of Various Emerging Nanoelectronics (RAVEN) program, contract FA8650-17-C-9113. G.B. also acknowledges financial support from the Intra-Create thematic grant NRF2019-THE002-0006 ``Retinal Analytics through Machine learning aiding Physics'' (RAMP) by Singapore's National Research Foundation. The opinions expressed herein are the authors' solely, and do not reflect the opinions of the sponsors. 

The authors disclose no conflicts of interest.
\end{acknowledgments}

\appendix

\section{\label{app:2DLSE} LSE as a composition of \\ 2D Fourier transforms}
In this section, we derive Eq.~(\ref{eq:LSE_2D_FT}). Fourier transforming $\psi-\psi_0$ yields
\begin{align*} \label{eq:2DLSE}
    &\hat{\mathcal{F}}_{xy}
    \left[\psi(\bm{r}) - \psi_0(\bm{r})\right] (k_x, k_y, z) \\
    &= \int dx\,dy\, \ee{-ik_xx-ik_yy} \left[\psi(\bm{r}) - \psi_0(\bm{r})\right] \\
    &= \int dx\,dy\, \ee{-ik_xx-ik_yy} \int d\bm{r}^\prime \,\, G(\bm{r}-\bm{r}^\prime) 
    V(\bm{r}^\prime)  \psi(\bm{r}^\prime).
    \stepcounter{equation}\tag{\theequation}
\end{align*}
Using the Weyl expansion, Eq.~(\ref{eq:weyl_expansion}), the Green's function can also be expressed as a 2D Fourier transform. Then we obtain
\begin{align*} \label{eq:2DLSE_2}
    &\hat{\mathcal{F}}_{xy}
    \left[\psi(\bm{r}) - \psi_0(\bm{r})\right] (k_x, k_y, z) \\
    &=\frac{i}{8\pi^2} \int dx\,dy
    \int d\bm{r}^\prime 
    \int dk_x^\prime dk_y^\prime \,\, 
    \ee{-ik_xx-ik_yy} \\
    & \qquad \times 
    \frac{ \ee{i\left(k_x^\prime (x-x^\prime) + k_y^\prime (y-y^\prime) + k_z^\prime |z-z^\prime| \right)} }{k_z^\prime}
    V(\bm{r}^\prime)  \psi(\bm{r}^\prime) \\
    &= \frac{i}{8\pi^2} \int d\bm{r}^\prime 
    V(\bm{r}^\prime)  \psi(\bm{r}^\prime)
    \int dk_x^\prime dk_y^\prime \\
    & \times 
    \frac{ \ee{-i\left(k_x^\prime x^\prime + k_y^\prime y^\prime - k_z^\prime |z-z^\prime| \right)} }{k_z^\prime}
    \underbrace{\int dx\,dy \,\,
    \ee{i\left(x(k_x^\prime-k_x) + y(k_y^\prime-k_y) \right)}}_{(2\pi)^2 \delta(k_x-k_x^\prime) \delta(k_y-k_y^\prime)} \\
    &= \frac{i}{2} \int d\bm{r}^\prime 
    V(\bm{r}^\prime)  \psi(\bm{r}^\prime)\,\,
    \frac{ \ee{-i\left(k_x x^\prime + k_y y^\prime - k_z |z-z^\prime| \right)} }{k_z} \\
    &= \frac{i}{2} \int dz^\prime \,\,
    \frac{ \ee{i k_z |z-z^\prime| } }{k_z}
    \int dx^\prime dy^\prime
    V(\bm{r}^\prime)  \psi(\bm{r}^\prime) \,\,
    \ee{-i\left(k_x x^\prime + k_y y^\prime \right)}\\
    &= \frac{i}{2} \int dz^\prime \,\,
    \frac{ \ee{i k_z |z-z^\prime| } }{k_z}
    \hat{\mathcal{F}}_{xy} 
    \left[V(\bm{r})  \psi(\bm{r})\right] (k_x, k_y, z^\prime).
    \stepcounter{equation}\tag{\theequation}
\end{align*}
Taking the inverse Fourier transform in Eq.~(\ref{eq:2DLSE_2}) finalizes the derivation leading to Eq.~(\ref{eq:LSE_2D_FT}).

\section{\label{app:FDTD} Comparison between FDTD and LSE}
\begin{figure}
\centering
\includegraphics[width=\linewidth]{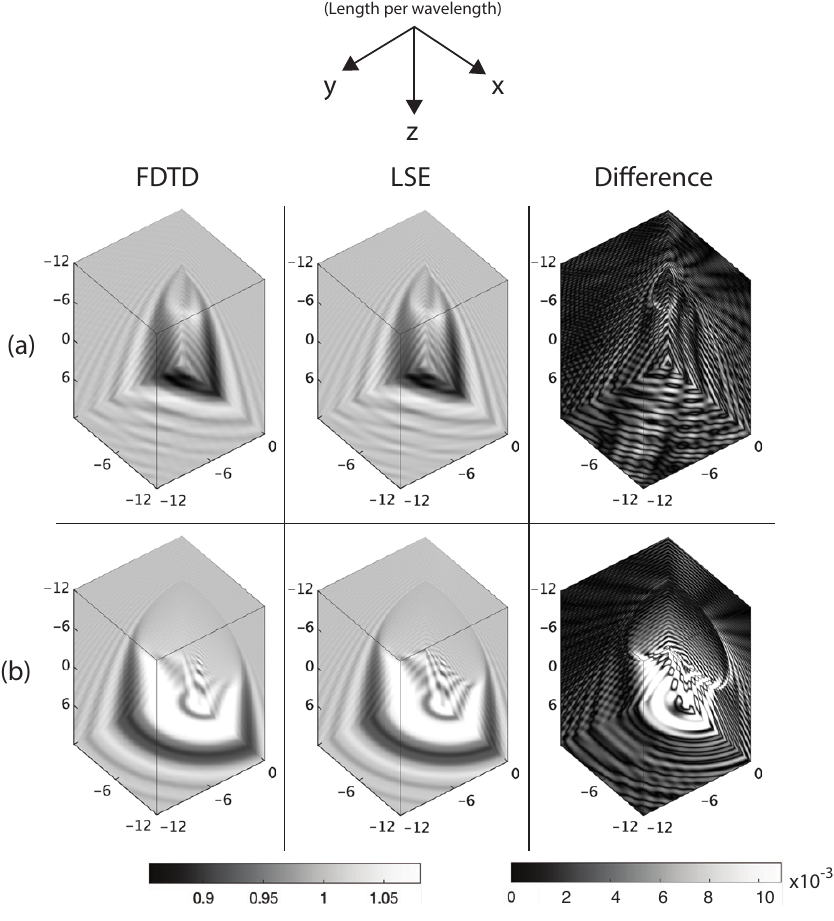}
\caption{\label{fig:lumerical} Comparison of scattered fields from FDTD and LSE. Two different potentials are considered where the mean refractive index is 1.02. These potentials are marked with (a) and (b). Difference refers to the elementwise absolute error divided by the maximum field amplitude.}
\end{figure}

To test the estimation quality of LSE, we compare it with FDTD solutions from the Lumerical \cite{lumerical} 3D Electromagnetic Simulator. In Fig.~(\ref{fig:lumerical}), it can be observed that the high frequency interference patterns are approximated well by the LSE. The numerical difference in each voxel is less than one percent of the maximum amplitude value. In Table~\ref{tab:lumerical}, we list quantitative results considering six different potentials. These results further corroborate the validity of the LSE.

\begin{table}[t]
\caption{\label{tab:lumerical} Image quality metrics on fields from LSE and FDTD. We consider 6 different potentials which consist of spheres. The mean refractive index of spherical potentials is $1.02$. The size of a cubic computational box is $24\lambda$. The phase is unwrapped along the optical axis. The full width at half maximum of the Gaussian window in SSIM is $\lambda/2$.}
\begin{ruledtabular}
\begin{tabular}{cccc}
&
\textrm{SSIM}&
\textrm{PSNR}&
\textrm{Relative $L_1$}\\
\colrule
Amplitude & $0.982$ & $42.162$ & $3.592\times 10^{-3}$\\
Phase & $0.999$ & $42.465$ & $2.486\times 10^{-2}$\\
\end{tabular}
\end{ruledtabular}
\end{table}

\section{\label{app:BornConv} Potential bound for \\ convergence of the Born series}

Previous studies discuss how to estimate the operator norm of the LSE integral operator and thus guarantee the convergence of the Born series. For example, \cite{manning1965error} requires
\begin{equation} \label{eq:manning1}
    2\int \max_{\theta,\phi} \left| V(r,\theta,\phi) \right| r dr < 1,
\end{equation}
where $r$, $\theta$, and $\phi$ are radial distance, polar angle, and azimuthal angle in the spherical coordinate system. Considering the simplest case, let us assume a Mie scattering condition in which a sphere of radius $R$ scatters a plane wave. Then Eq.~(\ref{eq:manning1}) becomes
\begin{equation}
    \left(\frac{n}{\nb}\right)^2 < 1 + \frac{1}{(\nb k_0R)^2}.
\end{equation}
Similarly, \cite{kilgore2017convergence} suggests
\begin{equation} \label{eq:kilgore1}
    \left(\frac{n}{\nb}\right)^2 < 1+
    \frac{1}{17/2 (\nb k_0R)^2 + 2\sqrt{74}(\nb k_0R) + 105}.
\end{equation}
By comparison, our discussion in Sec.~\ref{sec:convBorn} concludes that it is sufficient to satisfy
\begin{equation} \label{eq:oursbornconv1}
    \left(\frac{n}{\nb}\right)^2 < 
    1 + \frac{1}{2\sqrt{3}(\nb k_0R)}
\end{equation}
to make the Born series convergent. The scalar wave approximation already requires $\nb k_0R \gg 1$, which means that $(\nb k_0R)^2$ terms in Eqs.~(\ref{eq:manning1})-(\ref{eq:kilgore1}) increase quickly. This makes the estimation on the upper bound of $n$ too close to 1. On the contrary, Eq.~(\ref{eq:oursbornconv1}) shows the first-order dependency on $\nb k_0R$, which relaxes the requirement on $n$.

\begin{table}[t]
\caption{\label{tab:BPM_12} Image quality metrics on fields from BPM with $\xi=1$ and $\xi=2$ when the size $L$ of a cubic computational box changes. We consider 15 different potentials which consist of spheres. The mean refractive index of spherical potentials is $1.02$. The phase is unwrapped along the optical axis. The full width at half maximum of the Gaussian window in SSIM is $\lambda/2$.}
\begin{ruledtabular}
\begin{tabular}{cccc}
&
\textrm{SSIM}&
\textrm{PSNR}&
\textrm{Relative $L_1$}\\
\colrule
$L=16\lambda$, amplitude & $1.000$ & $64.461$ & $2.911\times 10^{-4}$\\
$L=24\lambda$, amplitude & $1.000$ & $62.679$ & $3.636\times 10^{-4}$\\
$L=32\lambda$, amplitude & $1.000$ & $62.884$ & $4.319\times 10^{-4}$\\
$L=40\lambda$, amplitude & $1.000$ & $62.849$ & $4.999\times 10^{-4}$\\
$L=16\lambda$, phase & $1.000$ & $91.573$ & $1.944\times 10^{-5}$\\
$L=24\lambda$, phase & $1.000$ & $91.587$ & $1.882\times 10^{-5}$\\
$L=32\lambda$, phase & $1.000$ & $91.601$ & $1.841\times 10^{-5}$\\
$L=40\lambda$, phase & $1.000$ & $91.416$ & $1.813\times 10^{-5}$\\
\end{tabular}
\end{ruledtabular}
\end{table}

\begin{table}[t!]
\caption{\label{tab:1_08_scale} Image quality metrics on fields from LSE and BPM when the size $L$ of a cubic computational box changes. We consider 15 different potentials which consist of spheres. The mean refractive index of spherical potentials is $1.08$. The phase is unwrapped along the optical axis. The full width at half maximum of the Gaussian window in SSIM is $\lambda/2$.}
\begin{ruledtabular}
\begin{tabular}{cccc}
&
\textrm{SSIM}&
\textrm{PSNR}&
\textrm{Relative $L_1$}\\
\colrule
$L=16\lambda$, amplitude & $0.923$ & $36.243$ & $3.392\times 10^{-2}$\\
$L=24\lambda$, amplitude & $0.930$ & $37.200$ & $4.170\times 10^{-2}$\\
$L=32\lambda$, amplitude & $0.932$ & $37.691$ & $4.932\times 10^{-2}$\\
$L=40\lambda$, amplitude & $0.937$ & $38.330$ & $5.528\times 10^{-2}$\\
$L=16\lambda$, phase & $0.989$ & $38.068$ & $4.121\times 10^{-2}$\\
$L=24\lambda$, phase & $0.990$ & $40.656$ & $2.769\times 10^{-2}$\\
$L=32\lambda$, phase & $0.986$ & $41.105$ & $2.220\times 10^{-2}$\\
$L=40\lambda$, phase & $0.983$ & $40.212$ & $1.938\times 10^{-2}$\\
\end{tabular}
\end{ruledtabular}
\end{table}

\section{Numerical comparison on different $\xi$ \\ in BPM's wave modulation \label{sec:xi}}
\begin{figure*}
\centering
\includegraphics[width=\textwidth]{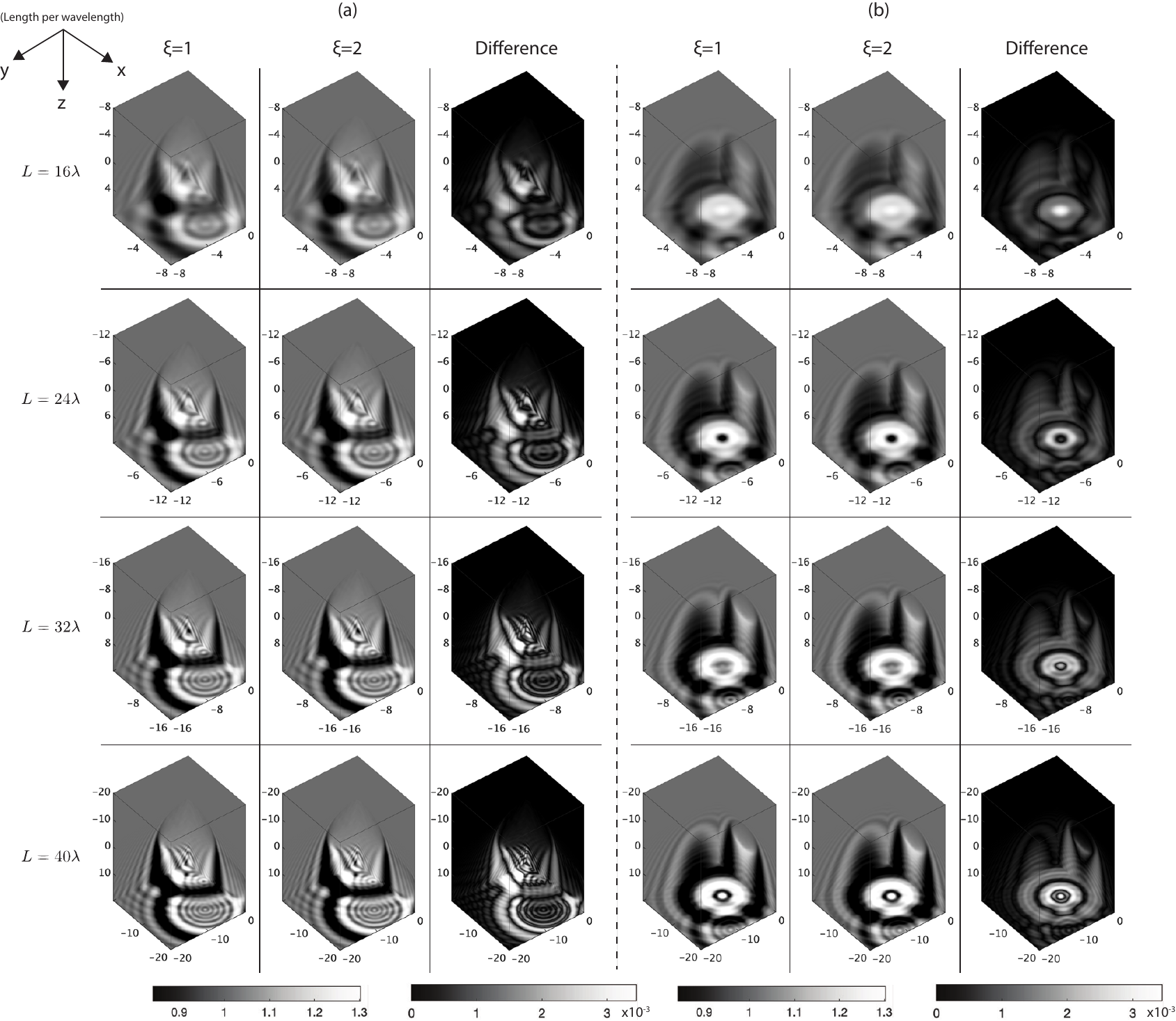}
\caption{\label{fig:xi_comparison} Scattered fields estimated from BPM with different $\xi$ choices: $\xi=1$ and $\xi=2$. We consider potentials which consist of spheres. The size $L$ of a cubic computational box is changed from $16\lambda$ to $40\lambda$. The mean refractive index of spherical potentials is $1.02$. We show two different objects, which are marked with (a) and (b). Difference refers to the elementwise absolute error divided by the maximum field amplitude.}
\end{figure*}

Based on the discussion in Sec.~\ref{sec:bpm_xi_value}, we compare field estimations from different $\xi$ in BPM. In Fig.~\ref{fig:xi_comparison}, it is shown that there is no significant difference in scattered amplitudes and the elementwise difference is less than one percent of maximum amplitude value. This can be quantitatively validated in Table~\ref{tab:BPM_12} where SSIM and PSNR exhibit very high values. Hence, we may conclude that $\xi=1$ and $\xi=2$ in the phase modulation term would not significantly influence the field estimation, except some unusual cases.

\section{Supplement to size dependence of \\ error between LSE and BPM \label{sec:LSE_BPM_addition}}
\begin{figure*}
\centering
\includegraphics[width=\textwidth]{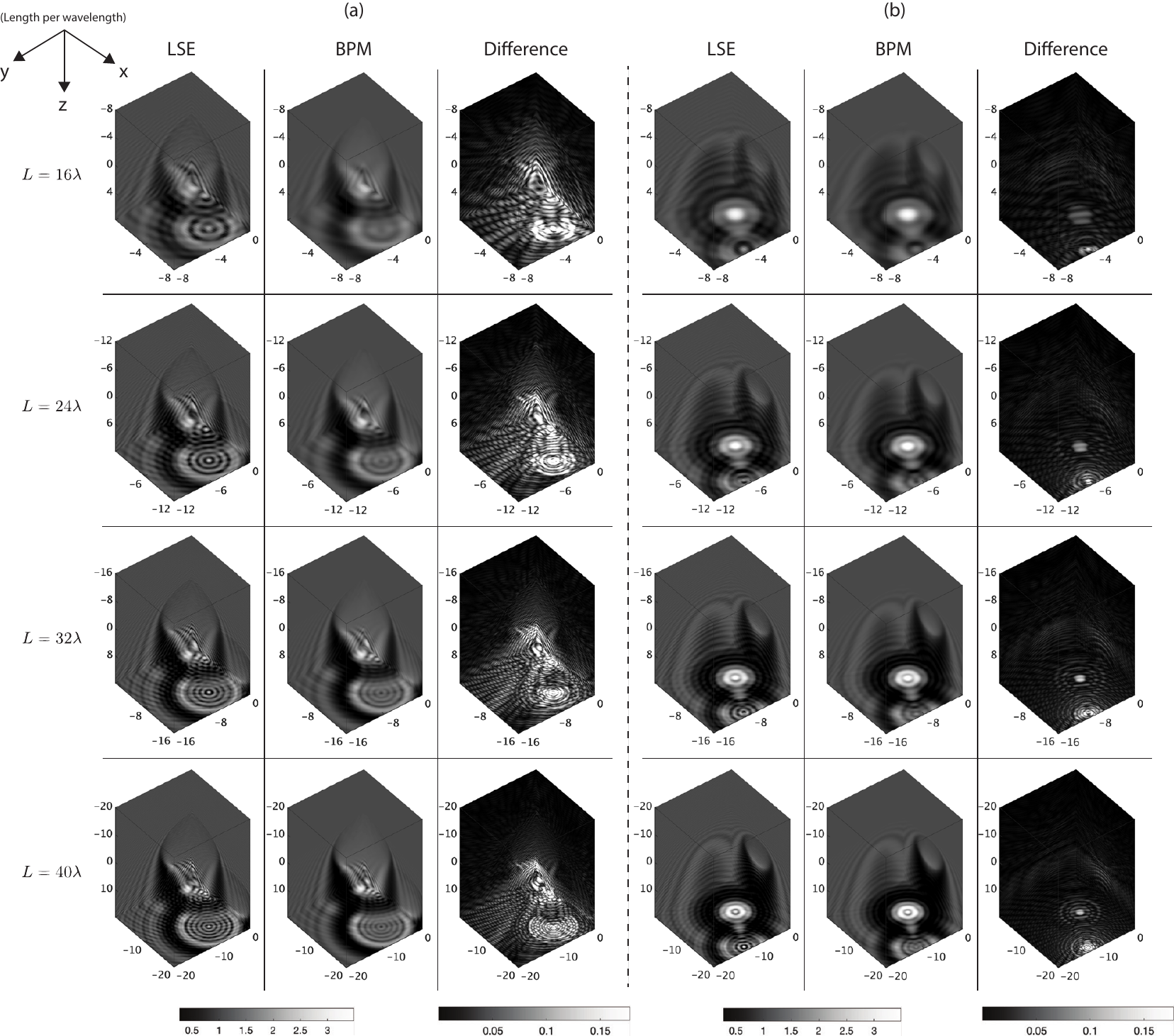}
\caption{\label{fig:index_1_08} Scattered fields estimated from LSE and BPM when the size $L$ of a cubic computational box changes. We consider potentials which consist of spheres. The mean refractive index of spherical potentials is $1.08$. We show two different objects, which are marked with (a) and (b). Difference refers to the elementwise absolute error divided by the maximum field amplitude.}
\end{figure*}

Corroborating results in Fig.~(\ref{fig:index_1_02}) and Table~\ref{tab:1_02_scale}, we conduct additional experiments on the size dependency of the error between LSE and BPM under a higher refractive index $n$ condition. Specifically, we set $n=1.08$. In Fig.~(\ref{fig:index_1_08}), we can observe the expected tendency of BPM to well approximate interference patterns of LSE as size increases, except at strong focal points. Table~\ref{tab:1_08_scale} lists corresponding quantitative results, which show decrease in SSIM and PSNR for the phase from large potentials. This may be attributed to the increased ill-conditionedness of the LSE operator \cite{zepeda2016fast} and fine oscillatory features, which reduces the numerical stability of the simulation.

\bibliography{ref}

\begin{thebibliography}{32}%
\makeatletter
\providecommand \@ifxundefined [1]{%
 \@ifx{#1\undefined}
}%
\providecommand \@ifnum [1]{%
 \ifnum #1\expandafter \@firstoftwo
 \else \expandafter \@secondoftwo
 \fi
}%
\providecommand \@ifx [1]{%
 \ifx #1\expandafter \@firstoftwo
 \else \expandafter \@secondoftwo
 \fi
}%
\providecommand \natexlab [1]{#1}%
\providecommand \enquote  [1]{``#1''}%
\providecommand \bibnamefont  [1]{#1}%
\providecommand \bibfnamefont [1]{#1}%
\providecommand \citenamefont [1]{#1}%
\providecommand \href@noop [0]{\@secondoftwo}%
\providecommand \href [0]{\begingroup \@sanitize@url \@href}%
\providecommand \@href[1]{\@@startlink{#1}\@@href}%
\providecommand \@@href[1]{\endgroup#1\@@endlink}%
\providecommand \@sanitize@url [0]{\catcode `\\12\catcode `\$12\catcode
  `\&12\catcode `\#12\catcode `\^12\catcode `\_12\catcode `\%12\relax}%
\providecommand \@@startlink[1]{}%
\providecommand \@@endlink[0]{}%
\providecommand \url  [0]{\begingroup\@sanitize@url \@url }%
\providecommand \@url [1]{\endgroup\@href {#1}{\urlprefix }}%
\providecommand \urlprefix  [0]{URL }%
\providecommand \Eprint [0]{\href }%
\providecommand \doibase [0]{https://doi.org/}%
\providecommand \selectlanguage [0]{\@gobble}%
\providecommand \bibinfo  [0]{\@secondoftwo}%
\providecommand \bibfield  [0]{\@secondoftwo}%
\providecommand \translation [1]{[#1]}%
\providecommand \BibitemOpen [0]{}%
\providecommand \bibitemStop [0]{}%
\providecommand \bibitemNoStop [0]{.\EOS\space}%
\providecommand \EOS [0]{\spacefactor3000\relax}%
\providecommand \BibitemShut  [1]{\csname bibitem#1\endcsname}%
\let\auto@bib@innerbib\@empty
\bibitem [{\citenamefont {Paganin}\ \emph {et~al.}(2006)\citenamefont {Paganin}
  \emph {et~al.}}]{paganin2006coherent}%
  \BibitemOpen
  \bibfield  {author} {\bibinfo {author} {\bibfnamefont {D.}~\bibnamefont
  {Paganin}} \emph {et~al.},\ }\href@noop {} {\emph {\bibinfo {title} {Coherent
  X-ray optics}}},\ \bibinfo {number} {6}\ (\bibinfo  {publisher} {Oxford
  University Press on Demand},\ \bibinfo {year} {2006})\BibitemShut {NoStop}%
\bibitem [{\citenamefont {Marks}(2006)}]{marks2006family}%
  \BibitemOpen
  \bibfield  {author} {\bibinfo {author} {\bibfnamefont {D.~L.}\ \bibnamefont
  {Marks}},\ }\bibfield  {title} {\bibinfo {title} {A family of approximations
  spanning the {Born} and {Rytov} scattering series},\ }\href@noop {}
  {\bibfield  {journal} {\bibinfo  {journal} {Optics express}\ }\textbf
  {\bibinfo {volume} {14}},\ \bibinfo {pages} {8837} (\bibinfo {year}
  {2006})}\BibitemShut {NoStop}%
\bibitem [{\citenamefont {Pham}\ \emph {et~al.}(2020)\citenamefont {Pham},
  \citenamefont {Soubies}, \citenamefont {Ayoub}, \citenamefont {Lim},
  \citenamefont {Psaltis},\ and\ \citenamefont {Unser}}]{pham2020three}%
  \BibitemOpen
  \bibfield  {author} {\bibinfo {author} {\bibfnamefont {T.-A.}\ \bibnamefont
  {Pham}}, \bibinfo {author} {\bibfnamefont {E.}~\bibnamefont {Soubies}},
  \bibinfo {author} {\bibfnamefont {A.}~\bibnamefont {Ayoub}}, \bibinfo
  {author} {\bibfnamefont {J.}~\bibnamefont {Lim}}, \bibinfo {author}
  {\bibfnamefont {D.}~\bibnamefont {Psaltis}},\ and\ \bibinfo {author}
  {\bibfnamefont {M.}~\bibnamefont {Unser}},\ }\bibfield  {title} {\bibinfo
  {title} {Three-dimensional optical diffraction tomography with
  lippmann-schwinger model},\ }\href@noop {} {\bibfield  {journal} {\bibinfo
  {journal} {IEEE Transactions on Computational Imaging}\ }\textbf {\bibinfo
  {volume} {6}},\ \bibinfo {pages} {727} (\bibinfo {year} {2020})}\BibitemShut
  {NoStop}%
\bibitem [{\citenamefont {B{\"u}rgel}\ \emph {et~al.}(2017)\citenamefont
  {B{\"u}rgel}, \citenamefont {Kazimierski},\ and\ \citenamefont
  {Lechleiter}}]{burgel2017sparsity}%
  \BibitemOpen
  \bibfield  {author} {\bibinfo {author} {\bibfnamefont {F.}~\bibnamefont
  {B{\"u}rgel}}, \bibinfo {author} {\bibfnamefont {K.~S.}\ \bibnamefont
  {Kazimierski}},\ and\ \bibinfo {author} {\bibfnamefont {A.}~\bibnamefont
  {Lechleiter}},\ }\bibfield  {title} {\bibinfo {title} {A sparsity
  regularization and total variation based computational framework for the
  inverse medium problem in scattering},\ }\href@noop {} {\bibfield  {journal}
  {\bibinfo  {journal} {Journal of Computational Physics}\ }\textbf {\bibinfo
  {volume} {339}},\ \bibinfo {pages} {1} (\bibinfo {year} {2017})}\BibitemShut
  {NoStop}%
\bibitem [{\citenamefont {Liu}\ \emph {et~al.}(2017)\citenamefont {Liu},
  \citenamefont {Liu}, \citenamefont {Mansour}, \citenamefont {Boufounos},
  \citenamefont {Waller},\ and\ \citenamefont {Kamilov}}]{liu2017seagle}%
  \BibitemOpen
  \bibfield  {author} {\bibinfo {author} {\bibfnamefont {H.-Y.}\ \bibnamefont
  {Liu}}, \bibinfo {author} {\bibfnamefont {D.}~\bibnamefont {Liu}}, \bibinfo
  {author} {\bibfnamefont {H.}~\bibnamefont {Mansour}}, \bibinfo {author}
  {\bibfnamefont {P.~T.}\ \bibnamefont {Boufounos}}, \bibinfo {author}
  {\bibfnamefont {L.}~\bibnamefont {Waller}},\ and\ \bibinfo {author}
  {\bibfnamefont {U.~S.}\ \bibnamefont {Kamilov}},\ }\bibfield  {title}
  {\bibinfo {title} {Seagle: Sparsity-driven image reconstruction under
  multiple scattering},\ }\href@noop {} {\bibfield  {journal} {\bibinfo
  {journal} {IEEE Transactions on Computational Imaging}\ }\textbf {\bibinfo
  {volume} {4}},\ \bibinfo {pages} {73} (\bibinfo {year} {2017})}\BibitemShut
  {NoStop}%
\bibitem [{\citenamefont {Kamilov}\ \emph {et~al.}(2015)\citenamefont
  {Kamilov}, \citenamefont {Papadopoulos}, \citenamefont {Shoreh},
  \citenamefont {Goy}, \citenamefont {Vonesch}, \citenamefont {Unser},\ and\
  \citenamefont {Psaltis}}]{kamilov2015learning}%
  \BibitemOpen
  \bibfield  {author} {\bibinfo {author} {\bibfnamefont {U.~S.}\ \bibnamefont
  {Kamilov}}, \bibinfo {author} {\bibfnamefont {I.~N.}\ \bibnamefont
  {Papadopoulos}}, \bibinfo {author} {\bibfnamefont {M.~H.}\ \bibnamefont
  {Shoreh}}, \bibinfo {author} {\bibfnamefont {A.}~\bibnamefont {Goy}},
  \bibinfo {author} {\bibfnamefont {C.}~\bibnamefont {Vonesch}}, \bibinfo
  {author} {\bibfnamefont {M.}~\bibnamefont {Unser}},\ and\ \bibinfo {author}
  {\bibfnamefont {D.}~\bibnamefont {Psaltis}},\ }\bibfield  {title} {\bibinfo
  {title} {Learning approach to optical tomography},\ }\href@noop {} {\bibfield
   {journal} {\bibinfo  {journal} {Optica}\ }\textbf {\bibinfo {volume} {2}},\
  \bibinfo {pages} {517} (\bibinfo {year} {2015})}\BibitemShut {NoStop}%
\bibitem [{\citenamefont {Kamilov}\ \emph {et~al.}(2016)\citenamefont
  {Kamilov}, \citenamefont {Papadopoulos}, \citenamefont {Shoreh},
  \citenamefont {Goy}, \citenamefont {Vonesch}, \citenamefont {Unser},\ and\
  \citenamefont {Psaltis}}]{kamilov2016optical}%
  \BibitemOpen
  \bibfield  {author} {\bibinfo {author} {\bibfnamefont {U.~S.}\ \bibnamefont
  {Kamilov}}, \bibinfo {author} {\bibfnamefont {I.~N.}\ \bibnamefont
  {Papadopoulos}}, \bibinfo {author} {\bibfnamefont {M.~H.}\ \bibnamefont
  {Shoreh}}, \bibinfo {author} {\bibfnamefont {A.}~\bibnamefont {Goy}},
  \bibinfo {author} {\bibfnamefont {C.}~\bibnamefont {Vonesch}}, \bibinfo
  {author} {\bibfnamefont {M.}~\bibnamefont {Unser}},\ and\ \bibinfo {author}
  {\bibfnamefont {D.}~\bibnamefont {Psaltis}},\ }\bibfield  {title} {\bibinfo
  {title} {Optical tomographic image reconstruction based on beam propagation
  and sparse regularization},\ }\href@noop {} {\bibfield  {journal} {\bibinfo
  {journal} {IEEE Transactions on Computational Imaging}\ }\textbf {\bibinfo
  {volume} {2}},\ \bibinfo {pages} {59} (\bibinfo {year} {2016})}\BibitemShut
  {NoStop}%
\bibitem [{\citenamefont {Goy}\ \emph {et~al.}(2019)\citenamefont {Goy},
  \citenamefont {Rughoobur}, \citenamefont {Li}, \citenamefont {Arthur},
  \citenamefont {Akinwande},\ and\ \citenamefont {Barbastathis}}]{goy2019high}%
  \BibitemOpen
  \bibfield  {author} {\bibinfo {author} {\bibfnamefont {A.}~\bibnamefont
  {Goy}}, \bibinfo {author} {\bibfnamefont {G.}~\bibnamefont {Rughoobur}},
  \bibinfo {author} {\bibfnamefont {S.}~\bibnamefont {Li}}, \bibinfo {author}
  {\bibfnamefont {K.}~\bibnamefont {Arthur}}, \bibinfo {author} {\bibfnamefont
  {A.~I.}\ \bibnamefont {Akinwande}},\ and\ \bibinfo {author} {\bibfnamefont
  {G.}~\bibnamefont {Barbastathis}},\ }\bibfield  {title} {\bibinfo {title}
  {High-resolution limited-angle phase tomography of dense layered objects
  using deep neural networks},\ }\href@noop {} {\bibfield  {journal} {\bibinfo
  {journal} {Proceedings of the National Academy of Sciences}\ }\textbf
  {\bibinfo {volume} {116}},\ \bibinfo {pages} {19848} (\bibinfo {year}
  {2019})}\BibitemShut {NoStop}%
\bibitem [{\citenamefont {Osnabrugge}\ \emph {et~al.}(2016)\citenamefont
  {Osnabrugge}, \citenamefont {Leedumrongwatthanakun},\ and\ \citenamefont
  {Vellekoop}}]{osnabrugge2016convergent}%
  \BibitemOpen
  \bibfield  {author} {\bibinfo {author} {\bibfnamefont {G.}~\bibnamefont
  {Osnabrugge}}, \bibinfo {author} {\bibfnamefont {S.}~\bibnamefont
  {Leedumrongwatthanakun}},\ and\ \bibinfo {author} {\bibfnamefont {I.~M.}\
  \bibnamefont {Vellekoop}},\ }\bibfield  {title} {\bibinfo {title} {A
  convergent {Born} series for solving the inhomogeneous helmholtz equation in
  arbitrarily large media},\ }\href@noop {} {\bibfield  {journal} {\bibinfo
  {journal} {Journal of computational physics}\ }\textbf {\bibinfo {volume}
  {322}},\ \bibinfo {pages} {113} (\bibinfo {year} {2016})}\BibitemShut
  {NoStop}%
\bibitem [{\citenamefont {Tahir}\ \emph {et~al.}(2019)\citenamefont {Tahir},
  \citenamefont {Kamilov},\ and\ \citenamefont {Tian}}]{tahir2019holographic}%
  \BibitemOpen
  \bibfield  {author} {\bibinfo {author} {\bibfnamefont {W.}~\bibnamefont
  {Tahir}}, \bibinfo {author} {\bibfnamefont {U.~S.}\ \bibnamefont {Kamilov}},\
  and\ \bibinfo {author} {\bibfnamefont {L.}~\bibnamefont {Tian}},\ }\bibfield
  {title} {\bibinfo {title} {Holographic particle localization under multiple
  scattering},\ }\href@noop {} {\bibfield  {journal} {\bibinfo  {journal}
  {Advanced Photonics}\ }\textbf {\bibinfo {volume} {1}},\ \bibinfo {pages}
  {036003} (\bibinfo {year} {2019})}\BibitemShut {NoStop}%
\bibitem [{\citenamefont {Kr{\"u}ger}\ \emph {et~al.}(2017)\citenamefont
  {Kr{\"u}ger}, \citenamefont {Brenner},\ and\ \citenamefont
  {Kienle}}]{kruger2017solution}%
  \BibitemOpen
  \bibfield  {author} {\bibinfo {author} {\bibfnamefont {B.}~\bibnamefont
  {Kr{\"u}ger}}, \bibinfo {author} {\bibfnamefont {T.}~\bibnamefont
  {Brenner}},\ and\ \bibinfo {author} {\bibfnamefont {A.}~\bibnamefont
  {Kienle}},\ }\bibfield  {title} {\bibinfo {title} {Solution of the
  inhomogeneous maxwell’s equations using a {Born} series},\ }\href@noop {}
  {\bibfield  {journal} {\bibinfo  {journal} {Optics express}\ }\textbf
  {\bibinfo {volume} {25}},\ \bibinfo {pages} {25165} (\bibinfo {year}
  {2017})}\BibitemShut {NoStop}%
\bibitem [{\citenamefont {Hohage}(2006)}]{hohage2006fast}%
  \BibitemOpen
  \bibfield  {author} {\bibinfo {author} {\bibfnamefont {T.}~\bibnamefont
  {Hohage}},\ }\bibfield  {title} {\bibinfo {title} {Fast numerical solution of
  the electromagnetic medium scattering problem and applications to the inverse
  problem},\ }\href@noop {} {\bibfield  {journal} {\bibinfo  {journal} {Journal
  of Computational Physics}\ }\textbf {\bibinfo {volume} {214}},\ \bibinfo
  {pages} {224} (\bibinfo {year} {2006})}\BibitemShut {NoStop}%
\bibitem [{\citenamefont {Feit}\ and\ \citenamefont
  {Fleck}(1988)}]{feit1988beam}%
  \BibitemOpen
  \bibfield  {author} {\bibinfo {author} {\bibfnamefont {M.}~\bibnamefont
  {Feit}}\ and\ \bibinfo {author} {\bibfnamefont {J.}~\bibnamefont {Fleck}},\
  }\bibfield  {title} {\bibinfo {title} {Beam nonparaxiality, filament
  formation, and beam breakup in the self-focusing of optical beams},\
  }\href@noop {} {\bibfield  {journal} {\bibinfo  {journal} {JOSA B}\ }\textbf
  {\bibinfo {volume} {5}},\ \bibinfo {pages} {633} (\bibinfo {year}
  {1988})}\BibitemShut {NoStop}%
\bibitem [{\citenamefont {Colton}\ \emph {et~al.}(1998)\citenamefont {Colton},
  \citenamefont {Kress},\ and\ \citenamefont {Kress}}]{colton1998inverse}%
  \BibitemOpen
  \bibfield  {author} {\bibinfo {author} {\bibfnamefont {D.~L.}\ \bibnamefont
  {Colton}}, \bibinfo {author} {\bibfnamefont {R.}~\bibnamefont {Kress}},\ and\
  \bibinfo {author} {\bibfnamefont {R.}~\bibnamefont {Kress}},\ }\href@noop {}
  {\emph {\bibinfo {title} {Inverse acoustic and electromagnetic scattering
  theory}}},\ Vol.~\bibinfo {volume} {93}\ (\bibinfo  {publisher} {Springer},\
  \bibinfo {year} {1998})\BibitemShut {NoStop}%
\bibitem [{\citenamefont {Lim}\ \emph {et~al.}(2019)\citenamefont {Lim},
  \citenamefont {Ayoub}, \citenamefont {Antoine},\ and\ \citenamefont
  {Psaltis}}]{lim2019high}%
  \BibitemOpen
  \bibfield  {author} {\bibinfo {author} {\bibfnamefont {J.}~\bibnamefont
  {Lim}}, \bibinfo {author} {\bibfnamefont {A.~B.}\ \bibnamefont {Ayoub}},
  \bibinfo {author} {\bibfnamefont {E.~E.}\ \bibnamefont {Antoine}},\ and\
  \bibinfo {author} {\bibfnamefont {D.}~\bibnamefont {Psaltis}},\ }\bibfield
  {title} {\bibinfo {title} {High-fidelity optical diffraction tomography of
  multiple scattering samples},\ }\href@noop {} {\bibfield  {journal} {\bibinfo
   {journal} {Light: Science \& Applications}\ }\textbf {\bibinfo {volume}
  {8}},\ \bibinfo {pages} {1} (\bibinfo {year} {2019})}\BibitemShut {NoStop}%
\bibitem [{\citenamefont {Chen}\ \emph {et~al.}(2020)\citenamefont {Chen},
  \citenamefont {Ren}, \citenamefont {Liu}, \citenamefont {Chowdhury},\ and\
  \citenamefont {Waller}}]{chen2020multi}%
  \BibitemOpen
  \bibfield  {author} {\bibinfo {author} {\bibfnamefont {M.}~\bibnamefont
  {Chen}}, \bibinfo {author} {\bibfnamefont {D.}~\bibnamefont {Ren}}, \bibinfo
  {author} {\bibfnamefont {H.-Y.}\ \bibnamefont {Liu}}, \bibinfo {author}
  {\bibfnamefont {S.}~\bibnamefont {Chowdhury}},\ and\ \bibinfo {author}
  {\bibfnamefont {L.}~\bibnamefont {Waller}},\ }\bibfield  {title} {\bibinfo
  {title} {Multi-layer {Born} multiple-scattering model for 3d phase
  microscopy},\ }\href@noop {} {\bibfield  {journal} {\bibinfo  {journal}
  {Optica}\ }\textbf {\bibinfo {volume} {7}},\ \bibinfo {pages} {394} (\bibinfo
  {year} {2020})}\BibitemShut {NoStop}%
\bibitem [{\citenamefont {Ying}(2015)}]{ying2015sparsifying}%
  \BibitemOpen
  \bibfield  {author} {\bibinfo {author} {\bibfnamefont {L.}~\bibnamefont
  {Ying}},\ }\bibfield  {title} {\bibinfo {title} {Sparsifying preconditioner
  for the lippmann--schwinger equation},\ }\href@noop {} {\bibfield  {journal}
  {\bibinfo  {journal} {Multiscale Modeling \& Simulation}\ }\textbf {\bibinfo
  {volume} {13}},\ \bibinfo {pages} {644} (\bibinfo {year} {2015})}\BibitemShut
  {NoStop}%
\bibitem [{\citenamefont {Brenner}\ and\ \citenamefont
  {Singer}(1993)}]{brenner1993light}%
  \BibitemOpen
  \bibfield  {author} {\bibinfo {author} {\bibfnamefont {K.-H.}\ \bibnamefont
  {Brenner}}\ and\ \bibinfo {author} {\bibfnamefont {W.}~\bibnamefont
  {Singer}},\ }\bibfield  {title} {\bibinfo {title} {Light propagation through
  microlenses: a new simulation method},\ }\href@noop {} {\bibfield  {journal}
  {\bibinfo  {journal} {Applied optics}\ }\textbf {\bibinfo {volume} {32}},\
  \bibinfo {pages} {4984} (\bibinfo {year} {1993})}\BibitemShut {NoStop}%
\bibitem [{\citenamefont {Schmalz}\ \emph {et~al.}(2010)\citenamefont
  {Schmalz}, \citenamefont {Schmalz}, \citenamefont {Gureyev},\ and\
  \citenamefont {Pavlov}}]{schmalz2010derivation}%
  \BibitemOpen
  \bibfield  {author} {\bibinfo {author} {\bibfnamefont {J.~A.}\ \bibnamefont
  {Schmalz}}, \bibinfo {author} {\bibfnamefont {G.}~\bibnamefont {Schmalz}},
  \bibinfo {author} {\bibfnamefont {T.~E.}\ \bibnamefont {Gureyev}},\ and\
  \bibinfo {author} {\bibfnamefont {K.~M.}\ \bibnamefont {Pavlov}},\ }\bibfield
   {title} {\bibinfo {title} {On the derivation of the green’s function for
  the helmholtz equation using generalized functions},\ }\href@noop {}
  {\bibfield  {journal} {\bibinfo  {journal} {American Journal of Physics}\
  }\textbf {\bibinfo {volume} {78}},\ \bibinfo {pages} {181} (\bibinfo {year}
  {2010})}\BibitemShut {NoStop}%
\bibitem [{\citenamefont {Born}\ and\ \citenamefont
  {Wolf}(2013)}]{born2013principles}%
  \BibitemOpen
  \bibfield  {author} {\bibinfo {author} {\bibfnamefont {M.}~\bibnamefont
  {Born}}\ and\ \bibinfo {author} {\bibfnamefont {E.}~\bibnamefont {Wolf}},\
  }\href@noop {} {\emph {\bibinfo {title} {Principles of optics:
  electromagnetic theory of propagation, interference and diffraction of
  light}}}\ (\bibinfo  {publisher} {Elsevier},\ \bibinfo {year}
  {2013})\BibitemShut {NoStop}%
\bibitem [{\citenamefont {Ishizuka}\ and\ \citenamefont
  {Uyeda}(1977)}]{ishizuka1977new}%
  \BibitemOpen
  \bibfield  {author} {\bibinfo {author} {\bibfnamefont {K.}~\bibnamefont
  {Ishizuka}}\ and\ \bibinfo {author} {\bibfnamefont {N.}~\bibnamefont
  {Uyeda}},\ }\bibfield  {title} {\bibinfo {title} {A new theoretical and
  practical approach to the multislice method},\ }\href@noop {} {\bibfield
  {journal} {\bibinfo  {journal} {Acta Crystallographica Section A: Crystal
  Physics, Diffraction, Theoretical and General Crystallography}\ }\textbf
  {\bibinfo {volume} {33}},\ \bibinfo {pages} {740} (\bibinfo {year}
  {1977})}\BibitemShut {NoStop}%
\bibitem [{\citenamefont {Manning}(1965)}]{manning1965error}%
  \BibitemOpen
  \bibfield  {author} {\bibinfo {author} {\bibfnamefont {I.}~\bibnamefont
  {Manning}},\ }\bibfield  {title} {\bibinfo {title} {Error and convergence
  bounds for the {B}orn expansion},\ }\href@noop {} {\bibfield  {journal}
  {\bibinfo  {journal} {Physical Review}\ }\textbf {\bibinfo {volume} {139}},\
  \bibinfo {pages} {B495} (\bibinfo {year} {1965})}\BibitemShut {NoStop}%
\bibitem [{\citenamefont {Natterer}(2004)}]{natterer2004error}%
  \BibitemOpen
  \bibfield  {author} {\bibinfo {author} {\bibfnamefont {F.}~\bibnamefont
  {Natterer}},\ }\bibfield  {title} {\bibinfo {title} {An error bound for the
  {Born} approximation},\ }\href@noop {} {\bibfield  {journal} {\bibinfo
  {journal} {Inverse problems}\ }\textbf {\bibinfo {volume} {20}},\ \bibinfo
  {pages} {447} (\bibinfo {year} {2004})}\BibitemShut {NoStop}%
\bibitem [{\citenamefont {Vico}\ \emph {et~al.}(2016)\citenamefont {Vico},
  \citenamefont {Greengard},\ and\ \citenamefont {Ferrando}}]{vico2016fast}%
  \BibitemOpen
  \bibfield  {author} {\bibinfo {author} {\bibfnamefont {F.}~\bibnamefont
  {Vico}}, \bibinfo {author} {\bibfnamefont {L.}~\bibnamefont {Greengard}},\
  and\ \bibinfo {author} {\bibfnamefont {M.}~\bibnamefont {Ferrando}},\
  }\bibfield  {title} {\bibinfo {title} {Fast convolution with free-space
  green's functions},\ }\href@noop {} {\bibfield  {journal} {\bibinfo
  {journal} {Journal of Computational Physics}\ }\textbf {\bibinfo {volume}
  {323}},\ \bibinfo {pages} {191} (\bibinfo {year} {2016})}\BibitemShut
  {NoStop}%
\bibitem [{\citenamefont {Kilgore}\ \emph {et~al.}(2017)\citenamefont
  {Kilgore}, \citenamefont {Moskow},\ and\ \citenamefont
  {Schotland}}]{kilgore2017convergence}%
  \BibitemOpen
  \bibfield  {author} {\bibinfo {author} {\bibfnamefont {K.}~\bibnamefont
  {Kilgore}}, \bibinfo {author} {\bibfnamefont {S.}~\bibnamefont {Moskow}},\
  and\ \bibinfo {author} {\bibfnamefont {J.~C.}\ \bibnamefont {Schotland}},\
  }\bibfield  {title} {\bibinfo {title} {Convergence of the {Born} and inverse
  {Born} series for electromagnetic scattering},\ }\href@noop {} {\bibfield
  {journal} {\bibinfo  {journal} {Applicable analysis}\ }\textbf {\bibinfo
  {volume} {96}},\ \bibinfo {pages} {1737} (\bibinfo {year}
  {2017})}\BibitemShut {NoStop}%
\bibitem [{\citenamefont {Teague}(1983)}]{teague1983deterministic}%
  \BibitemOpen
  \bibfield  {author} {\bibinfo {author} {\bibfnamefont {M.~R.}\ \bibnamefont
  {Teague}},\ }\bibfield  {title} {\bibinfo {title} {Deterministic phase
  retrieval: a green’s function solution},\ }\href@noop {} {\bibfield
  {journal} {\bibinfo  {journal} {JOSA}\ }\textbf {\bibinfo {volume} {73}},\
  \bibinfo {pages} {1434} (\bibinfo {year} {1983})}\BibitemShut {NoStop}%
\bibitem [{\citenamefont {Thomson}\ and\ \citenamefont
  {Chapman}(1983)}]{thomson1983wide}%
  \BibitemOpen
  \bibfield  {author} {\bibinfo {author} {\bibfnamefont {D.~J.}\ \bibnamefont
  {Thomson}}\ and\ \bibinfo {author} {\bibfnamefont {N.}~\bibnamefont
  {Chapman}},\ }\bibfield  {title} {\bibinfo {title} {A wide-angle split-step
  algorithm for the parabolic equation},\ }\href@noop {} {\bibfield  {journal}
  {\bibinfo  {journal} {The Journal of the Acoustical Society of America}\
  }\textbf {\bibinfo {volume} {74}},\ \bibinfo {pages} {1848} (\bibinfo {year}
  {1983})}\BibitemShut {NoStop}%
\bibitem [{\citenamefont {Feit}\ and\ \citenamefont
  {Fleck}(1978)}]{feit1978light}%
  \BibitemOpen
  \bibfield  {author} {\bibinfo {author} {\bibfnamefont {M.}~\bibnamefont
  {Feit}}\ and\ \bibinfo {author} {\bibfnamefont {J.}~\bibnamefont {Fleck}},\
  }\bibfield  {title} {\bibinfo {title} {Light propagation in graded-index
  optical fibers},\ }\href@noop {} {\bibfield  {journal} {\bibinfo  {journal}
  {Applied optics}\ }\textbf {\bibinfo {volume} {17}},\ \bibinfo {pages} {3990}
  (\bibinfo {year} {1978})}\BibitemShut {NoStop}%
\bibitem [{\citenamefont {Wang}\ \emph {et~al.}(2004)\citenamefont {Wang},
  \citenamefont {Bovik}, \citenamefont {Sheikh},\ and\ \citenamefont
  {Simoncelli}}]{wang2004image}%
  \BibitemOpen
  \bibfield  {author} {\bibinfo {author} {\bibfnamefont {Z.}~\bibnamefont
  {Wang}}, \bibinfo {author} {\bibfnamefont {A.~C.}\ \bibnamefont {Bovik}},
  \bibinfo {author} {\bibfnamefont {H.~R.}\ \bibnamefont {Sheikh}},\ and\
  \bibinfo {author} {\bibfnamefont {E.~P.}\ \bibnamefont {Simoncelli}},\
  }\bibfield  {title} {\bibinfo {title} {Image quality assessment: from error
  visibility to structural similarity},\ }\href@noop {} {\bibfield  {journal}
  {\bibinfo  {journal} {IEEE transactions on image processing}\ }\textbf
  {\bibinfo {volume} {13}},\ \bibinfo {pages} {600} (\bibinfo {year}
  {2004})}\BibitemShut {NoStop}%
\bibitem [{\citenamefont {Hore}\ and\ \citenamefont
  {Ziou}(2010)}]{hore2010image}%
  \BibitemOpen
  \bibfield  {author} {\bibinfo {author} {\bibfnamefont {A.}~\bibnamefont
  {Hore}}\ and\ \bibinfo {author} {\bibfnamefont {D.}~\bibnamefont {Ziou}},\
  }\bibfield  {title} {\bibinfo {title} {Image quality metrics: Psnr vs.
  ssim},\ }in\ \href@noop {} {\emph {\bibinfo {booktitle} {2010 20th
  international conference on pattern recognition}}}\ (\bibinfo {organization}
  {IEEE},\ \bibinfo {year} {2010})\ pp.\ \bibinfo {pages}
  {2366--2369}\BibitemShut {NoStop}%
\bibitem [{lum()}]{lumerical}%
  \BibitemOpen
  \href@noop {} {\bibinfo {title} {Lumerical inc.}},\ \bibinfo {howpublished}
  {\url{ https://www.lumerical.com/}}\BibitemShut {NoStop}%
\bibitem [{\citenamefont {Zepeda-N{\'u}nez}\ and\ \citenamefont
  {Zhao}(2016)}]{zepeda2016fast}%
  \BibitemOpen
  \bibfield  {author} {\bibinfo {author} {\bibfnamefont {L.}~\bibnamefont
  {Zepeda-N{\'u}nez}}\ and\ \bibinfo {author} {\bibfnamefont {H.}~\bibnamefont
  {Zhao}},\ }\bibfield  {title} {\bibinfo {title} {Fast alternating
  bidirectional preconditioner for the 2d high-frequency lippmann--schwinger
  equation},\ }\href@noop {} {\bibfield  {journal} {\bibinfo  {journal} {SIAM
  Journal on Scientific Computing}\ }\textbf {\bibinfo {volume} {38}},\
  \bibinfo {pages} {B866} (\bibinfo {year} {2016})}\BibitemShut {NoStop}%
\end{thebibliography}%

\end{document}